\documentclass[sigplan,10pt]{acmart}
\skip\footinsauthorsaddresses=3pt plus 1pt minus 1pt
\skip\footinscopyrightpermission=3pt plus 1pt minus 1pt

\usepackage{enumitem}
\usepackage{caption}
\usepackage[utf8]{inputenc}
\usepackage{textgreek}
\usepackage{subcaption}
\usepackage{algorithmicx}
\usepackage{algorithm}
\usepackage{algcompatible}

\AtBeginDocument{
  
}

\copyrightyear{2026}
\acmYear{2026}
\setcopyright{cc}
\setcctype{by}
\acmConference[SIGOPS ATC 2026]{The 2026 ACM SIGOPS Annual Technical Conference}{November 15--18, 2026}{Hong Kong, Hong Kong}
\acmBooktitle{The 2026 ACM SIGOPS Annual Technical Conference (SIGOPS ATC 2026), November 15--18, 2026, Hong Kong, Hong Kong}
\acmDOI{10.1145/3828523.3857017}
\acmISBN{979-8-4007-2818-1/2026/11}

\begin{document}

\title{gKSM: GPU Memory Deduplication at Driver Layer for Serverless LLM Inference}

\author{Shuheng Gao}
\affiliation{
  \institution{University of Science and Technology of China}
  \city{Hefei}
  \country{China}
}
\email{shhgao@mail.ustc.edu.cn}

\author{Fan Guo}
\authornote{Corresponding authors.}
\affiliation{
  \institution{MinJiang University}
  \city{Fuzhou}
  \country{China}
}
\email{lps56@mail.ustc.edu.cn}

\author{Yongkun Li}
\authornotemark[1]
\affiliation{
  \institution{University of Science and Technology of China}
  \city{Hefei}
  \country{China}
}
\email{ykli@ustc.edu.cn}

\author{Ziye Yue}
\affiliation{
  \institution{University of Science and Technology of China}
  \city{Hefei}
  \country{China}
}
\email{ustc\_yzyeah@mail.ustc.edu.cn}

\author{Yinlong Xu}
\affiliation{
  \institution{University of Science and Technology of China}
  \city{Hefei}
  \country{China}
}
\email{ylxu@ustc.edu.cn}

\author{John C.S. Lui}
\affiliation{
  \institution{The Chinese University of Hong Kong}
  \city{Hong Kong}
  \country{China}
}
\email{cslui@cse.cuhk.edu.hk}

\begin{abstract}
In serverless LLM inference, cloud providers commonly locate multiple instances on a single GPU to improve utilization. However, each instance independently maintains its model weights and runtime state, making duplicate GPU memory the primary bottleneck. Memory deduplication is a widely-used technique for eliminating duplicate data in CPU systems, but extending it to GPU memory faces fundamental challenges. Effective GPU memory deduplication must avoid redundant data movement between CPU and GPU while preserving write isolation for shared pages without native copy-on-write (COW) support.

This paper presents gKSM, a transparent GPU memory deduplication framework for serverless LLM inference. gKSM addresses these challenges through CPU-GPU collaborative hash management that offloads hash computation to GPU kernels, zero-copy model weight deduplication that eliminates duplication before model weight transfer, and a cross-privilege page fault handler that combines a lightweight kernel hook and CUDA Virtual Memory Management (VMM) remapping in the user space. gKSM also designs a stable page scanning algorithm and a reuse-aware placement policy in cluster scale. Experiments show that gKSM reduces GPU memory usage by 59.8\% while maintaining high throughput compared to ServerlessLLM. Under memory constrained workloads, gKSM improves aggregate throughput by 2.15$\times$ and supports five instances concurrently. In cluster level simulation, gKSM achieves a service rate of 99.97\%, compared to 78.2\% for SLInfer and 26.9\% for ServerlessLLM.

\end{abstract}

\ccsdesc[500]{Software and its engineering~Memory management}
\ccsdesc[300]{Computer systems organization~Cloud computing}

\keywords{GPU memory management, serverless, memory deduplication, LLM inference}

\maketitle
\pagestyle{empty}
\section{Introduction}

Large language models (LLMs) are increasingly deployed as managed inference services in public cloud and private platforms~\cite{servegen2025,serverlessllm2024,punica2024,deepspeed_fastgen2024}. In many enterprise scenarios, to satisfy service level objectives (SLOs) and deployment customization in vertical industries, enterprises often deploy private model instances in the cloud instead of using public inference services~\cite{googlecloud_modelgarden_2025,alibabacloud_pai_eas}. These private services have different SLOs and dynamic request rates, making statically reserved GPU resources inefficient and expensive~\cite{yu2022orca,patel2024splitwise,agrawal2024sarathi,servegen2025}. To address this problem, serverless inference~\cite{serverlessllm2024,optimus2024,torpor2025,infaas2021,alpaserve2023} has emerged as an attractive deployment, allowing users to scale private model instances on demand without managing and scheduling GPU resources. Such services have already been adopted by major cloud providers, including Google~\cite{googlecloud_modelgarden_2025} and Alibaba Cloud~\cite{alibabacloud_pai_eas}. However, LLM inference fundamentally differs from traditional serverless functions~\cite{serverlesswild2020,li2022tetris,wang2021peeking,faasnet2021}. To avoid expensive model reloads during invocation, each inference instance must retain large model weights in GPU memory~\cite{serverlessllm2024,flexgen2023}. As a result, serverless LLM inference is constrained by persistent GPU memory consumption~\cite{yan2025Understanding,serverlessllm2024}.

Since online inference workloads are highly dynamic and intense, providers commonly locate multiple serverless inference instances on a single GPU to share compute resources over time~\cite{alpaserve2023,prism2025,torpor2025,alibaba_cluster_trace2018}. Although colocation improves compute utilization, each instance independently maintains its model weights, KV cache, and runtime state~\cite{kwon2023efficient,h2o2023}, making GPU memory the primary bottleneck that limits the number of colocated instances. Moreover, GPU memory usage across colocated instances is often highly duplicate. In practical deployment, many instances serve the same popular models or fine-tuned models derived from the same base model~\cite{lora2022,qlora2023,peft_survey2024,punica2024,slora2024}. Even when instances use different inference frameworks~\cite{kwon2023efficient,sglang2024,tensorrtllm2024,deepspeed_inference2022,deepspeed_fastgen2024}, their model weights remain largely identical. This duplication severely constrains the serving capacity of the inference system.


Existing works have proposed model-aware scheduling strategies to mitigate GPU memory waste. ServerlessLLM exploits the locality of model weights to reduce cold start and migration overhead~\cite{serverlessllm2024}. Prism dynamically offloads instances by tracking the access frequency of different models~\cite{prism2025}. SLInfer consolidates instances of the same model onto shared nodes via live migration and merges small request batches into larger ones~\cite{slinfer2026}. These strategies improve resource elasticity and GPU utilization~\cite{distserve2024,muxserve2024}. However, they operate at the level of instance scheduling and batch aggregation rather than directly eliminating duplicate memory pages at the physical level. They cannot share memory between instances running on different inference frameworks or exploit partial weight duplication between similar models. Furthermore, live migration introduces tail latency while weakening instance isolation. As a result, these scheduling approaches do not fully resolve the GPU memory duplication problem.

The scale of duplication is substantial. To confirm this problem, we perform a trace driven simulation of an A100 GPU cluster running a serverless inference workload of Alibaba Cloud~\cite{yan2025Understanding,faasnet2021,alibaba_cluster_trace2018}. Our experiments show that ServerlessLLM and conventional load balancing strategies accumulate up to 2000~GB of duplicate GPU memory, corresponding to a 70\% duplication rate. Even SLInfer, which explicitly consolidates identical instances, still retains a duplication rate of more than 50\% (see Figure~\ref{fig:memory_duplication}). Memory deduplication is a widely used technique in CPU systems, such as Linux KSM~\cite{singleton2012}. KSM identifies identical memory pages by scanning and merging them into a single physical copy, relying on COW to maintain data consistency when modifications occur. A transparent page level deduplication technique on the GPU side could therefore effectively eliminate duplicate model weights and runtime data across colocated instances.

However, CPU memory deduplication cannot be directly transplanted to GPU environments. Three fundamental challenges arise from the architectural differences between the two domains. First, duplicate data reside on the GPU while the metadata required for page tracking and sharing coordination is managed on the CPU. This separation makes CPU-side hashing prohibitively expensive and rules out a purely GPU-side management approach. Second, model weights must be loaded from the host to GPU memory over the PCIe bus during instance initialization. A conventional pipeline that copies first and deduplicates afterward wastes PCIe bandwidth and temporarily occupies duplicate memory, which leads to potential out-of-memory (OOM) risk. Third, after duplicate pages are shared, write isolation must be enforced. CPU KSM relies on the kernel's native COW mechanism, but the GPU driver, which is only partially open-source with a closed user-mode component, provides no COW semantics for deduplicated virtual memory.

To address these challenges, we propose gKSM, a GPU memory deduplication framework operating at the CUDA driver layer. Mainstream inference frameworks are built on CUDA software stack, their GPU memory operations pass through CUDA driver APIs. These interfaces provide a common interception point where gKSM can track allocations and copies and manage physical page sharing while preserving application-visible virtual addresses. This enables transparent page-level deduplication across independent instances, including those running different inference frameworks, without modifying the frameworks.  Our major contributions are summarized as follows.

\begin{itemize}[leftmargin=6pt]
\item We design a CPU--GPU collaborative hash management for page scanning. gKSM computes page hashes in the GPU, manages metadata and hash buckets in the CPU, and verifies candidate pages by exact comparison.
\item We propose zero-copy model weight deduplication based on duplicate data detection. By hooking CUDA memory copy during model loading, gKSM detects duplicate pages before transfer and avoids redundant memory allocations.
\item We design COW across user space and kernel space for shared GPU pages. gKSM captures write faults in the UVM path, reports them to a user space handler, and uses CUDA VMM to create private copies.
\item We implement gKSM at the CUDA driver layer, incorporating a stable page scanning algorithm for dynamic GPU memory and a reuse-aware instance placement policy for cluster deployment. Evaluated on a single GPU and cluster workloads, gKSM reduces GPU memory usage by up to 59.8\% with negligible overhead, and improves throughput by up to 2.15$\times$ under memory constrained workloads.
\end{itemize}

\section{Background and Motivation}
\subsection{Serverless Inference Deployment}
Serverless inference allows users to scale private model instances on demand without managing and scheduling GPU resources. However, serverless LLM inference differs significantly from traditional serverless function computation~\cite{serverlessllm2024,slinfer2026,prism2025}. A LLM inference instance typically needs to load model weights ranging from several to tens of GB and maintain runtime states such as KV cache during request processing~\cite{kwon2023efficient,h2o2023,flexgen2023}. Even when the inference workload is very low, model weights persistently occupy a large amount of GPU memory to avoid the high latency overhead of frequent cold starts. As a result, although serverless inference offers a more flexible deployment approach, its resource utilization remains constrained by GPU memory occupied by model weights and the dynamic fluctuations of online requests~\cite{agrawal2024sarathi,patel2024splitwise}.

\begin{figure}[t]
\centering
\begin{minipage}{0.5\textwidth}
\centering
\begin{subfigure}{0.49\linewidth}
    \includegraphics[width=\linewidth]{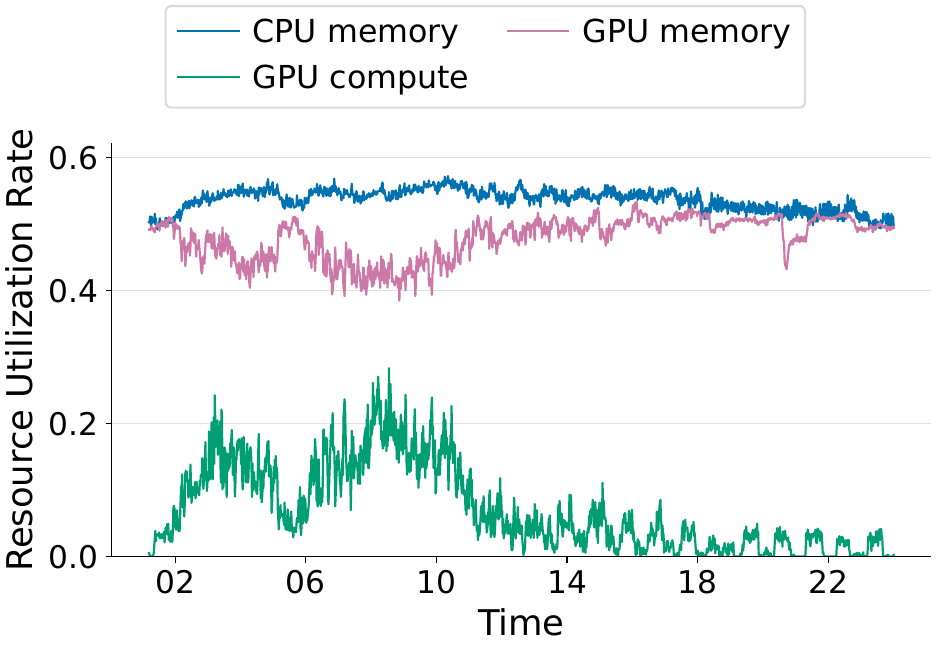}
    \caption{Average Resource Utilization}
    \label{fig:container_metrics_avg}
\end{subfigure}
\hfill
\begin{subfigure}{0.5\linewidth}
    \includegraphics[width=\linewidth]{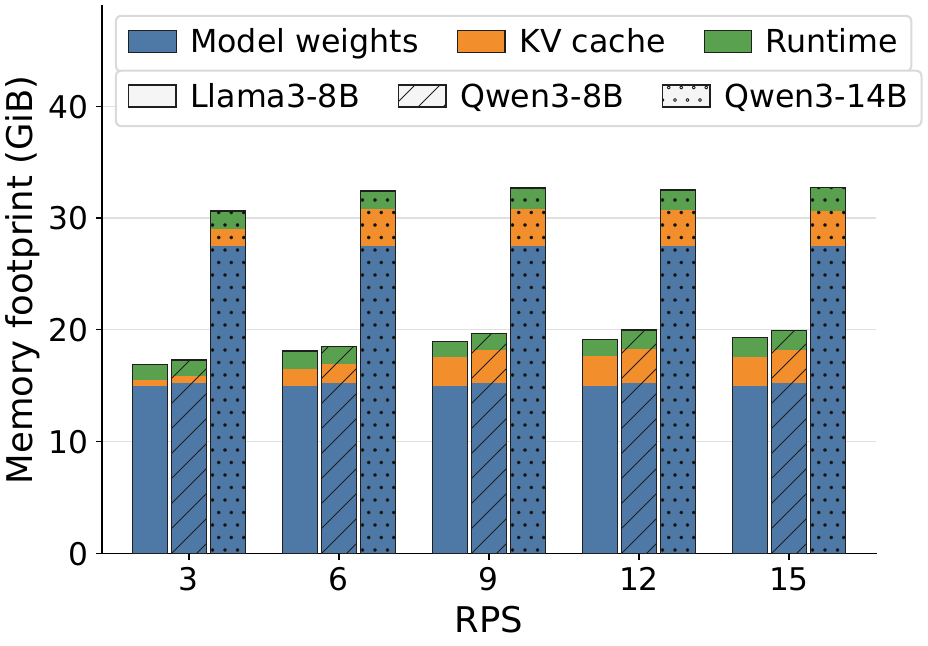}
    \caption{GPU Memory Breakdown}
    \label{fig:memory_breakdown}
\end{subfigure}
\end{minipage}
\caption{Resource utilization Analysis of the real world traces. Figure (a) illustrates the average resource utilization of serverless inference instances on Alibaba Cloud in one day. Figure (b) shows the inference memory breakdown for different models on A100 GPU.}
\label{fig:overall}
\end{figure}

To better understand this problem, we analyze a public serverless inference dataset from Alibaba Cloud~\cite{yan2025Understanding}. As shown in Figure~\ref{fig:container_metrics_avg}, for a single model instance, the average memory utilization is about 50\%, while the GPU computation utilization at peak is only around 20\%. This indicates that a single model instance often struggles to fully utilize the entire GPU. Due to the dynamic and intermittent nature of online inference requests, the arrival rate of the requests fluctuates over time, making it difficult for a model instance to maintain a high computational workload~\cite{yan2025Understanding,agrawal2024sarathi}. Under these circumstances, model weights persistently occupy GPU memory, while GPU compute units remain underutilized for extended periods, leading to a serious mismatch between memory occupation and computational resources. As shown in Figure~\ref{fig:memory_breakdown}, because most of the time instance workloads remain low, model weights dominate GPU memory usage. On the other hand, memory components such as KV cache and intermediate activations only account for less than 25\% of total memory usage~\cite{kwon2023efficient,h2o2023}.

Based on the above observations, cloud providers today typically adopt elastic resource sharing to improve GPU utilization~\cite{googlecloud_modelgarden_2025,alibabacloud_pai_eas,torpor2025}. For example, platforms such as Google Cloud and Alibaba Cloud support dynamic deployment and management of inference service instances on shared GPU resource pools~\cite{googlecloud_modelgarden_2025,alibabacloud_pai_eas}. Specifically, by deploying multiple model instances on a single GPU, the dynamic workload of different instances can complement each other over time~\cite{alpaserve2023,muxserve2024,llumnix2024}. When one model instance has idle computational resources, others can utilize them, thereby improving the overall GPU compute utilization. Compared to the deployment mode where a single instance exclusively occupies a GPU, deploying multiple instances per GPU enables finer-grained filling of GPU resource gaps, increasing both the number of model instances and the scale of online requests that a single GPU can support.

\subsection{Memory Duplication in Serverless Inference Systems}

Serverless deployment enables elastic scaling of inference instances across GPU resources. However, traditional inference instances adopt a static resource allocation model, where the system reserves fixed GPU memory and compute resources during instance startup and cannot dynamically adapt to fluctuations in the real-time request load~\cite{infaas2021,alpaserve2023}. To address this, existing works such as ServerlessLLM~\cite{serverlessllm2024}, Prism~\cite{prism2025} and SLInfer~\cite{slinfer2026} have proposed model-aware scheduling strategies~\cite{llumnix2024}. ServerlessLLM exploits the model weight locality to reduce cold start and migration overhead. Prism dynamically offloads instances by tracking the access frequency of different model instances in real time. SLInfer aggregates instances of the same model onto the same node via live migration and merges small batches into larger ones, addressing the memory redundancy caused by duplicate model weights. These strategies mitigate unnecessary GPU memory waste, enable elastic resource sharing, and improve GPU utilization~\cite{distserve2024,muxserve2024}.

\begin{figure}[t]
\centering
\hspace*{-0.3cm}
\begin{minipage}{0.5\textwidth}
\centering
\begin{subfigure}{0.45\linewidth}
    \raisebox{0.1cm}{
        \includegraphics[width=\linewidth]{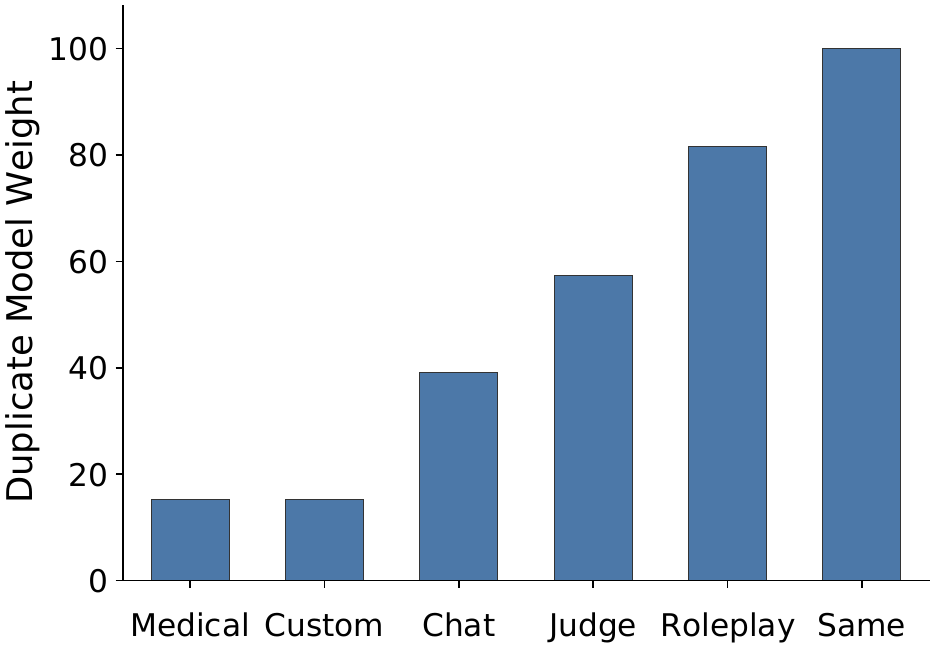}
    }
    \caption{Duplication of Fine-tuned Qwen3-8B from Huggingface}
    \label{fig:qwen3-8b-family}
\end{subfigure}
\hspace{0cm}
\begin{subfigure}{0.48\linewidth}
    \includegraphics[width=\linewidth]{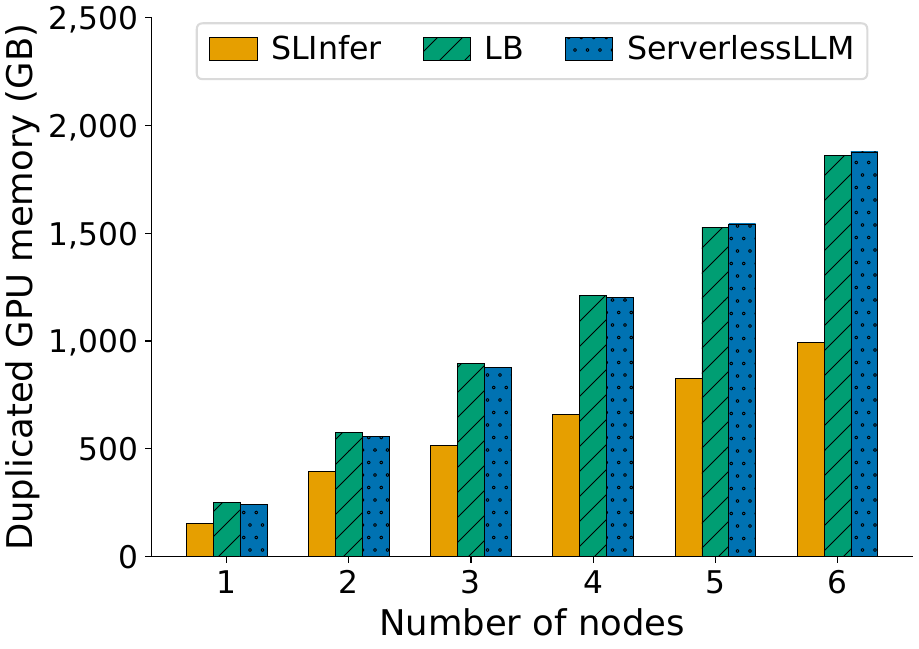}
    \caption{Duplication in GPU Cluster under different scheduling strategies}
    \label{fig:cluster_duplication}
\end{subfigure}
\end{minipage}
\caption{Duplicate GPU memory analysis. Figure (a) shows duplicate model weights between Qwen3-8B and its fine-tuned variants in Huggingface~\cite{huggingface_hub2022}. Figure (b) shows duplicate GPU memory under different scheduling strategies as the cluster size scales.}
\label{fig:memory_duplication}
\end{figure}

However, under existing model-aware scheduling strategies, inference clusters still face significant GPU memory duplication. To confirm this problem, we simulate and replay the cluster level inference process based on Alibaba Cloud inference traces~\cite{yan2025Understanding,faasnet2021,alibaba_cluster_trace2018} to quantify GPU memory duplication under different scheduling strategies. As shown in Figure~\ref{fig:cluster_duplication}, as the cluster size scales, the memory duplication of ServerlessLLM and traditional load balancing strategies reaches 2000~GB, with an overall duplication rate of 70\%, while SLInfer achieves a duplication rate of 50\%. Traditional load balancing strategies are model-agnostic, dispersing instances of the same model across different nodes and causing widespread memory duplication within the cluster. Although ServerlessLLM reduces cold start overhead by exploiting model weight locality, it tends to aggregate instances of the same model within GPU nodes. As a result, each GPU node still holds duplicate copies of model weights in its memory. SLInfer schedules instances with identical model weights to the same node and mitigates duplication by merging request batches, yet its scheduling strategy still faces fundamental limitations.

Existing scheduling strategies suffer from three intrinsic limitations. First, when the same model runs on different inference frameworks, such as vLLM~\cite{kwon2023efficient}, SGLang~\cite{sglang2024}, and TensorRT-LLM~\cite{tensorrtllm2024}, instances on different frameworks cannot share memory with each other. These frameworks differ fundamentally in request batching, prefill scheduling, and memory management, preventing cross-framework batch merging even when instances serve the same model. According to industry data, vLLM and TensorRT-LLM together dominate the NVIDIA inference ecosystem, with adoption rates of approximately 55\% and 45\% respectively, while SGLang is deployed on more than 300,000 GPUs~\cite{sglang2024}. Second, existing strategies only deduplicate the entirely identical model instances and cannot exploit partial weight sharing across different models. Many models are fine-tuned from the same base model~\cite{lora2022,qlora2023,peft_survey2024,punica2024,slora2024} or extend a shared language backbone with different vision encoders~\cite{llava2023,blip22023,qwenvl2023}. As shown in Figure~\ref{fig:qwen3-8b-family}, sharing the common base model of Qwen3-8B~\cite{qwen3_technical_report} saves 15.2\%--81.6\% GPU memory, yet existing schedulers cannot perceive such partial duplication. Third, live instance migration and batch merging~\cite{llumnix2024,distserve2024} introduce additional system overhead and undermine inter-tenant isolation. Live migration transfers instance context including runtime states and KV cache, incurring network overhead and service pauses that lead to tail latency~\cite{fastserve2023,clockwork2020}. Meanwhile, merging requests from different tenants into shared batches weakens the instance resource boundaries, causing performance interference under varying request patterns and latency SLOs~\cite{clockwork2020,missile2024}.

\subsection{Opportunities and Challenges of GPU Memory Deduplication}
Current approaches rely on the scheduling layer to continuously decide placement and migration, which fail to inherently eliminate duplicate memory footprints while incur substantial overhead in the scheduling and coordination layer. Memory deduplication~\cite{buddy2020, miller2012ksmpp,waldspurger2002memory,differenceengine2008,userguided2023} is an established technique that reduces physical memory usage by identifying and merging identical memory pages through scanning and comparison, while relying on COW to maintain data consistency when modifications occur. This approach has been widely implemented in CPU memory systems, such as Linux's KSM~\cite{singleton2012} and virtualization platforms~\cite{miller2012ksmpp,vmdedup2014,memorybuddies2009}.

In serverless inference scenarios, extending traditional CPU memory deduplication to the GPU offers an opportunity for transparent memory sharing across colocated instances. GPU memory deduplication at CUDA driver layer can provide an efficient and transparent solution that significantly reduces duplicate memory usage without affecting inference services. Existing mainstream inference frameworks, including vLLM and SGLang, are built on the CUDA software stack. Although they implement their own logical memory managers, their model weights, KV cache buffer, and other GPU runtime memory must ultimately be allocated and managed through CUDA driver APIs. This allows us to track all these allocations and memory copy operations without modifying any framework, preserving each allocation's virtual addresses while managing the physical backing pages. In contrast, runtime-level sharing is coupled to each framework's memory manager, making it less efficient and less transparent than driver-layer deduplication.

However, fundamental differences in memory management models, driver interfaces, and hardware capabilities between CPU and GPU architectures prevent direct transplantation of mature CPU deduplication mechanisms to GPU memory. The lack of native support for COW semantics and fine-grained page sharing in current GPU systems presents particular challenges~\cite{nvidia_cuda_vmm_2024,vattention2025,batchaware2020}. In general, these challenges can be summarized in the following three aspects:

\textbf{Challenge 1: Deduplication in heterogeneous memory.}
The lifecycle of GPU memory is managed from the CPU side, including physical memory allocation, virtual address remapping, access control and coordination. However, duplicate memory reside in the GPU side. Copying GPU pages back to host memory for CPU-side hashing introduces substantial data transfer overhead, and CPU hash computation throughput is much lower than what GPU parallelism can achieve. Therefore, metadata must be maintained on the CPU side, while hash computation and page comparison are performed on the GPU. However, moving the entire scanning and management logic to the GPU is infeasible because page metadata such as allocation handles, access permissions, and sharing state requires the logical control and data structure management capabilities of the CPU runtime. This physical separation between GPU-side data and CPU-side metadata creates a fundamental coordination challenge. Efficient duplicate detection requires a collaborative mechanism that leverages GPU parallelism for hash computation while retaining CPU-side metadata management, without incurring expensive data movement from GPU to CPU.

\textbf{Challenge 2: Eliminating redundant data transfer in the critical path.}
GPU inference workloads must first load model weights from CPU to GPU memory over the PCIe bus during instance initialization. A conventional deduplication pipeline would fully copy the model weights to the GPU, then scan and merge duplicate pages. This incurs unnecessary PCIe bandwidth consumption and temporarily occupies duplicate GPU memory before deduplication completes, leading to significant memory spike. Modern LLM model weights range from several to tens of gigabytes, and hundreds of instances may load overlapping weights concurrently at cluster scale, making these overheads significant. The challenge is to detect duplicates before GPU transfer to avoid redundant PCIe traffic and temporary duplicate memory usage.

\textbf{Challenge 3: Preserving write isolation without native COW support.}
After GPU memory pages are shared, write isolation must be preserved. In CPU-side KSM, this is handled by the kernel's native COW mechanism. When a process writes to a shared page, the CPU MMU triggers a page fault, and the kernel's fault handler transparently allocates a private copy and updates the page table, all within the kernel's privileged execution context. However, the GPU driver is only partially open-source. While the kernel-mode driver is open, the user-mode driver remains closed-source and neither layer exposes COW semantics for deduplicated GPU virtual memory. CUDA VMM provides page-level access control but lacks automatic COW services. Directly modifying the GPU kernel driver to implement full COW semantics would introduce significant engineering complexity and risk system stability across driver updates. Furthermore, the UVM kernel module manages page faults within its own kernel-space logic and does not coordinate with user-space CUDA VMM APIs for page remapping. The challenge is to implement COW semantics for shared GPU pages while keeping kernel modifications minimal and preserving system stability.

\begin{figure}[t]
  \centering
  \includegraphics[width=0.45\textwidth,height=0.22\textwidth]{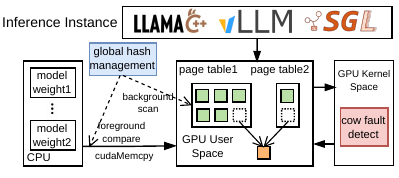}
  \caption{System design and architecture overview of gKSM}
  \label{fig:system}
\end{figure}

\section{System Design}
\subsection{Overall Architecture of gKSM}
We propose gKSM, a GPU memory deduplication system operating at the CUDA driver layer. Compared to traditional CPU memory deduplication such as KSM, gKSM addresses three fundamental challenges. First, GPU memory on the device and deduplication metadata on the host leave CPU hashing limited by data transfer and computation overhead, while GPU metadata management is limited by CPU coordination. gKSM addresses this through \textbf{CPU--GPU Collaborative Hash Management} (Section~\ref{subsec:cpu_gpu_hash}), which offloads hash computation to GPU kernels while retaining metadata control on the CPU. Second, a conventional pipeline that copies model weights first and deduplicates afterward wastes bandwidth and temporarily occupies duplicate memory. gKSM eliminates this through \textbf{Zero-Copy Model Weight Deduplication} (Section~\ref{subsec:zero_copy}), which intercepts the copy path and performs sharing inline with data transfer. Third, after sharing, write isolation requires COW support. Unlike KSM, which relies on the kernel's native COW mechanism, the partially open-source GPU driver provides no such COW service, and invasive driver modifications threaten system stability. gKSM proposes a \textbf{Cross-Privilege Page Fault Handler} (Section~\ref{subsec:cow_handler}), which combines a lightweight kernel hook with user-space CUDA VMM remapping to implement COW without invasive driver changes.

\begin{figure}[t]
  \centering
  \includegraphics[width=0.45\textwidth,height=0.22\textwidth]{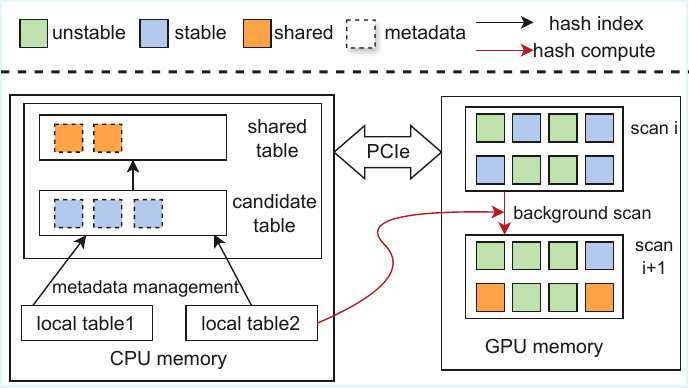}
  \caption{CPU--GPU collaborative hash management.}
  \label{fig:hash}
\end{figure}

\begin{figure}[t]
  \centering
  \includegraphics[width=\linewidth]{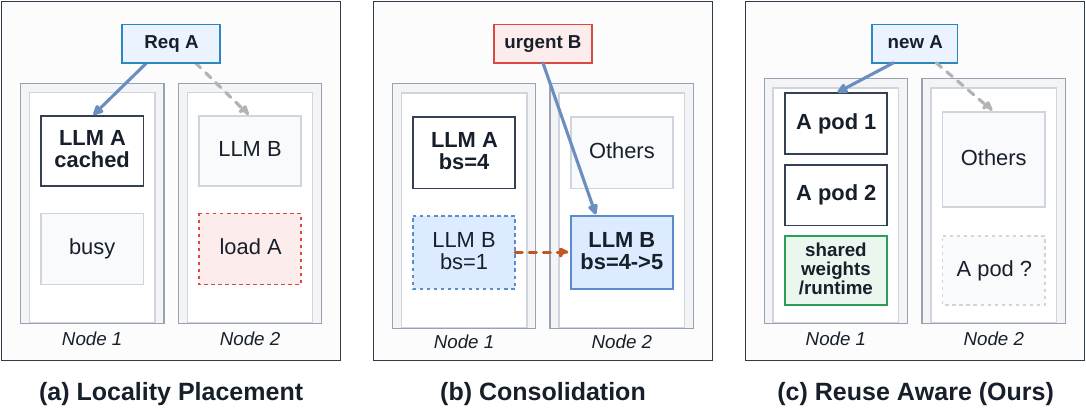}
  \caption{Different Scheduling Strategies}
  \label{fig:scheduler_strategies}
\end{figure}

As illustrated in Figure~\ref{fig:system}, gKSM implements this design through a transparent deduplication framework and a reuse-aware instance placement policy. The framework realizes GPU-accelerated stable page scanning that periodically identifies sharing opportunities in dynamic KV cache and activation pages~\cite{h2o2023,promptcache2024}, and transparent GPU page sharing via CUDA VMM that remaps duplicate pages to shared physical handles while preserving instance virtual address spaces. The placement scheduler decomposes each instance's memory into a shareable portion and a private portion, then selects the GPU with the lowest incremental memory cost under compute constraints. Figure~\ref{fig:scheduler_strategies} contrasts this policy with ServerlessLLM and SLInfer. Instead of relying on live migration or instance merging, gKSM exposes reuse opportunities directly to the memory layer. Since driver-layer deduplication requires reusable pages to reside on the same GPU~\cite{muxserve2024}, the placement policy locates instances that benefit from memory sharing. The workflow first deduplicates model parameters during initialization, then the background scanner identifies stable pages, while the COW handler preserves write isolation at runtime. The following subsections describe each module in detail.

\subsection{CPU--GPU Collaborative Hash Management}
\label{subsec:cpu_gpu_hash}
To coordinate the parallel computing capability of the GPU with the GPU memory management capability of the CPU, gKSM introduces a CPU-GPU collaborative hash management mechanism, as illustrated in Figure~\ref{fig:hash}. It offloads hash computation to the GPU while retaining metadata coordination on the CPU, leveraging GPU parallelism for batch hash computation and CPU control for metadata table management and candidate lookup.

On the GPU side, gKSM implements a CUDA kernel version of the xxHash algorithm~\cite{collet_xxhash_2024} to compute page hashes in batch. Hash matches identify candidate duplicates rather than directly authorizing page sharing. gKSM performs exact content comparison before merging, allowing us to use a fast non-cryptographic hash for candidate lookup. As shown in Table~\ref{tab:hash}, xxHash3 on the CPU side achieves only 9.92~GB/s, while xxHash3 on the GPU side reaches 167.70~GB/s, a 16.9$\times$ improvement. Using cryptographic algorithms on the CPU is even slower, such as SHA256~\cite{gilbert2003security}. 

\begin{table}[t]
  \centering
  \caption{The page hash computation overhead on NVIDIA A100 GPU. We measure the latency and throughput of different hash algorithms on both GPU and CPU.}
  \label{tab:hash}
  \begin{tabular}{@{}lcc@{}}
    \toprule
    Hash Algorithm & Latency(ms) & Throughput(GB/s) \\
    \midrule
    \texttt{xxHash3(GPU)}  & 47.7      & 167.70 \\
    \texttt{xxHash3(CPU)}  & 806.8     & 9.92   \\
    \texttt{SHA256(CPU)}   & 6166.3    & 1.31   \\
    \bottomrule
  \end{tabular}
\end{table}

On the CPU side, gKSM maintains a per-GPU metadata table that stores information including the virtual address, access permission, and hash value for each managed page. To coordinate deduplication across instances, gKSM further maintains a global shared table organized as hash buckets. This global table, implemented as an indexed array, supports efficient lookups and updates. A hash lookup returns candidate shared pages that share the same hash value. It is important to note that the above implementation still carries the risk of hash collisions. Therefore, a hash match does not directly lead to page sharing. Instead, it triggers an exact comparison of the page contents. Only when the page contents are completely identical will the current page be remapped to an existing shared page. If a hash collision occurs, the current page remains independent. By combining GPU-accelerated hash computation with optimized CPU-side metadata management, gKSM addresses the challenges of computational bottlenecks and cross-processor coordination, enabling fast and scalable page comparisons across instances.

\subsection{Zero-Copy Model Weight Deduplication}
\label{subsec:zero_copy}

As discussed in Section 2.3, a conventional pipeline that copies data first and deduplicates afterward wastes PCIe bandwidth and temporarily occupies duplicate GPU memory. These overheads are significant in the cluster scale, where hundreds of instances may load overlapping model weights concurrently, causing transient memory spikes and increasing the risk of OOM errors. To address this, gKSM introduces a zero-copy deduplication method integrated into the memory copy path. The core idea is to intercept the model weight transfer before it reaches GPU memory, so that duplicate data are never physically copied. 

Concretely, gKSM inserts a hook into the \texttt{cudaMemcpy()} path from CPU to GPU. Since model weights already reside in CPU memory before transfer, gKSM performs page hashing, candidate lookup, and exact content verification all on the CPU side. gKSM computes batched hashes in the hook path and uses them to query the global shared page table for candidate pages with matching hashes. If a match is found and the exact content comparison with the candidate pages retained pinned CPU buffer confirms this duplication, the virtual address of the current page is directly remapped to the physical handle of the existing shared page. The originally allocated physical memory is released immediately, without data ever being written there. If no match is found, the page is copied to the GPU as usual, and the new page is registered in the shared table and pinned buffer for future deduplication.

This approach ensures that only one physical copy of each unique model weight page is retained on the GPU, regardless of how many instances load the same model. To avoid unnecessary hash computation for model weight pages and other already-deduplicated data during subsequent background scans, these pages are marked with a flag so that they are excluded from further scanning. By avoiding redundant PCIe transfers and immediately releasing duplicate GPU allocations, zero-copy model weight deduplication achieves efficient memory savings with negligible impact on model initialization latency.


\subsection{Cross-Privilege Page Fault Handler}
\label{subsec:cow_handler}

After GPU memory deduplication, gKSM must preserve write isolation for shared GPU pages mapped by multiple instances. When one instance writes to the shared page, it must receive a private copy without affecting the other instances that are still reading from the original~\cite{batchaware2020}.

\begin{figure}[t]
  \centering
  \includegraphics[width=\linewidth]{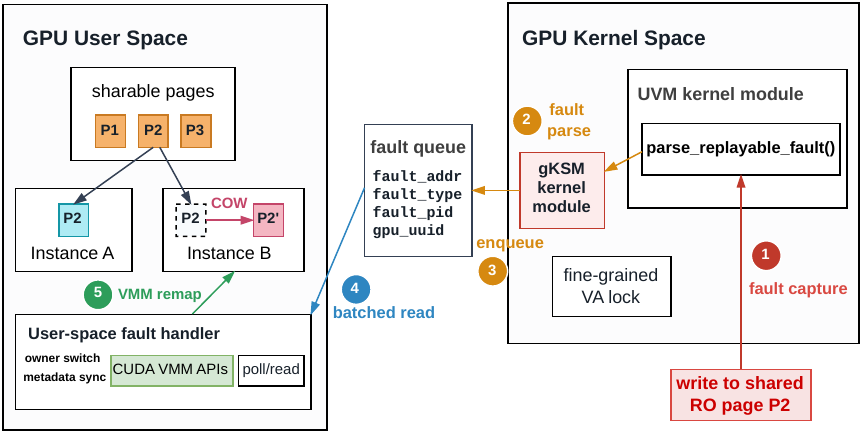}
  \caption{Cross-privilege COW fault handling for shared GPU pages. gKSM captures page faults in the kernel and remaps the fault page through the user-space CUDA VMM handler.}
  \label{fig:fault_handler}
\end{figure}

In traditional CPU memory deduplication, the kernel's native COW mechanism handles write isolation transparently within the privileged kernel context. However, the GPU environment lacks this native support. The partially open-source GPU driver provides no COW semantics. CUDA VMM lacks automatic page copy and remap services, and the UVM kernel module does not coordinate with user-space VMM APIs. To address this while keeping kernel modifications minimal, gKSM proposes a cross-privilege page fault handler that splits the COW workload across kernel space and user space. As shown in Figure~\ref{fig:fault_handler}, the kernel realizes a lightweight fault detection and reporting, while the user space handles page copying and virtual address remapping through CUDA VMM APIs.

After deduplication, gKSM maps all shared pages as read-only. When a GPU kernel attempts to write to a shared page, the access triggers a GPU page fault. On the kernel side, gKSM extends the UVM replayable fault path with a lightweight hook. For each GPU page fault, the hook checks whether the fault address falls within the gKSM-managed memory range and whether the fault is caused by a write access to a read-only shared page. For each captured fault, the kernel module records a fault descriptor that contains the fault address, fault type, instance identifier, and device identifier, then inserts it into a waiting queue exposed through a character device. The kernel does not perform expensive page copying or CUDA VMM remapping. Instead, it only reports fault events to the user space. The fault queue is protected by locks with a narrow scope, which reduces lock contention and keeps the GPU driver path simple and stable.

On the user side, a dedicated COW handler thread periodically fetches fault records from the waiting queue in batch. For each fault, the handler first validates the current local and global metadata. If the faulting instance is not the owner of the shared handle, the COW handler creates a private copy of the fault page for this instance using CUDA VMM allocation and memory copy. It then redirects the virtual address to the private copy via \texttt{cuMemMap()} and makes the new mapping writable via \texttt{cuMemSetAccess()}. The local metadata is updated to record the page as a private page, and the global shared table removes this reference from the sharer list of the original page. A COW fault is triggered only on the first write to a shared page. After a COW process is finished, subsequent writes to this private page follow the normal access pattern. 

If the faulting instance is the current owner of a shared page, the shared handle cannot be modified in place because other instances may still map to it. In this case, the COW handler first selects another instance from the sharer list as the new owner of the shared page group, preferring an instance on the same GPU. The ownership of the shared handle is then transferred to the new owner, and the faulting former owner is allocated a new private page. Its virtual address is remapped to this private copy with write access. If the reference count of the shared page becomes zero after all sharers have detached their references, the page is no longer shared. The COW handler restores write access for the remaining instance and removes the corresponding entry from the global shared table. During COW operations, gKSM protects the metadata of the faulting page with a shared memory mutex to prevent concurrent modification of the same shared page by multiple fault handlers. This design keeps the minimal kernel modification. Only a fault report hook is added to the existing UVM replayable fault path. All page copying, ownership management, and remapping logic resides in the user space, where CUDA VMM APIs provide a stable and well supported interface.

\begin{figure}[t]
  \centering
  \includegraphics[width=\linewidth]{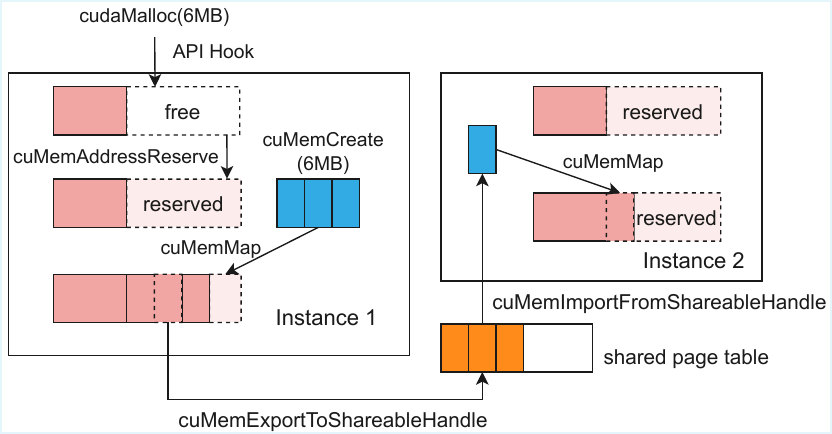}
  \caption{Page-level GPU memory management and sharing via CUDA VMM API.}
  \label{fig:vmm}
\end{figure}

\subsection{Implementation}
\subsubsection{Transparent GPU Page Sharing}
\label{subsec:gpu_page_sharing}
Once duplicate pages are identified, gKSM must share the underlying physical memory across instances while preserving each instance's virtual address space. In CPU-side KSM, the kernel directly modifies page table entries to share physical frames. The GPU environment lacks such direct page table manipulation, and conventional CUDA interfaces do not allow independent remapping or sharing of physical pages across instances. gKSM therefore implements transparent GPU page sharing on top of CUDA VMM APIs~\cite{vattention2025}, which expose fine-grained control over GPU virtual memory. These APIs support creating shareable physical handles, importing them into different address spaces, and independently controlling access permissions. gKSM leverages these primitives to turn duplicate pages into a single shareable physical handle while keeping each instance's virtual address unchanged. 

Concretely, gKSM intercepts GPU memory allocation requests and manages them at page granularity. For each managed page, gKSM maintains metadata to track its ownership, sharing state, and page permissions. Our implementation uses 2~MB as the deduplication granularity~\cite{mosaic2017}, which matches the allocation granularity reported by CUDA VMM on our platform. This granularity is also well suited for LLM inference workloads, since model weights are typically stored as large contiguous tensors with stable memory layouts across instances. As a result, most duplicate model weight regions can already be captured at 2~MB granularity, while using smaller pages would significantly increase metadata management, hashing, and remapping overhead.

As shown in Figure~\ref{fig:vmm}, the sharing process works as follows. Duplicate pages are identified either through zero-copy interception during model loading or through the background stable page scanner. Once identified, gKSM exports a shareable physical handle from the source instance using \texttt{cuMemExportToShareableHandle()}. This handle is then imported into the target instance's virtual address space, and the target's virtual address range is remapped to the imported physical handle via \texttt{cuMemMap()}. The target instance's original duplicate allocation is then released. Throughout this process, the virtual addresses of both instances remain unchanged. Inference kernels continue to access memory at the same addresses without any modification. After remapping, all shared pages are marked read-only via \texttt{cuMemSetAccess()} to preserve isolation across instances. This sets up the necessary condition for the COW fault handler described in Section~\ref{subsec:cow_handler}.

During normal inference execution, instances continue accessing GPU memory through their original virtual addresses. All reads are served directly from local HBM without remote memory access. gKSM only adds lightweight metadata management and occasional remap and permission updates during page sharing. In practice, these VMM operations occur infrequently and are concentrated mostly during initialization and deduplication events. They introduce negligible overhead and impact to the subsequent inference execution. For example, even the most expensive VMM operations, such as \texttt{cuMemExportToShareableHandle()} and \texttt{cuMemSetAccess()}, take only 100~$\mu$s and 296~$\mu$s on average in our implementation.

\subsubsection{Stable Page Scanning}
\label{subsec:stable_page}

Beyond model weights, LLM inference continuously generates dynamic pages such as KV cache and intermediate activations. These pages may contain significant duplication when instances serve similar requests or share prefix prompts~\cite{promptcache2024}, but unlike read-only model weights, they are subject to frequent modifications. Indiscriminately sharing them would lead to excessive COW faults, as a page written immediately after sharing must be privately reallocated and copied. In the GPU context, where COW involves a cross-privilege round trip (Section~\ref{subsec:cow_handler}), the cost of premature sharing is significantly higher than in CPU-side KSM. gKSM must therefore selectively identify pages that are sufficiently stable to benefit from deduplication.

To address this, gKSM introduces a stable page scanning mechanism that employs a lightweight background thread to periodically scan dynamic pages such as KV cache and intermediate activations. The core idea is to track page stability through repeated hash observations. If a page's hash value remains unchanged for $N$ consecutive scans, it is classified as stable and becomes eligible for deduplication. Pages whose hashes change between scans are considered unstable and are excluded from sharing, avoiding unnecessary COW overhead on transient data. The scanning interval follows an exponential-backoff policy: if a scan finds no deduplication opportunity, the interval doubles up to a configured maximum. Successful deduplication resets it to the base interval. The maximum interval controls the tradeoff between scanning overhead and responsiveness to new sharing opportunities: a smaller cap revisits dynamic pages more frequently, while a larger cap reduces repeated scanning when few stable duplicates are available. The background scanner leverages the same CPU--GPU collaborative hash computation mechanism described in Section~\ref{subsec:cpu_gpu_hash}, while the GPU computes page hashes and the CPU manages the global shared table.

Specifically, the GPU computes the current hash values for each page. If the hash value differs from the previous scan, the page is considered unstable and its stability counter is reset. If the hash remains consistent for $N$ consecutive scans, the page is deemed stable and gKSM proceeds to look up the corresponding hash bucket in the global candidate table. For each candidate page $q$ in the bucket, gKSM recomputes its hash value to filter out pages that may have been modified after being inserted into the candidate table. If $q$ resides in another instance, gKSM requests a hash recomputation on the remote instance. A matching hash alone does not guarantee page sharing due to the possibility of hash collisions. Before merging, gKSM performs an exact content comparison between the local page and the candidate. The two pages are remapped to the same physical page only if their contents are fully identical. If the comparison fails, gKSM treats the case as a hash collision or a stale candidate and continues to check the remaining candidates. If no candidate passes verification, the local page remains independent and may be inserted into the candidate table for future matching. By combining GPU-parallel hashing with stability-gated deduplication, gKSM effectively identifies sharing opportunities in transient data while minimizing scanning and COW overhead.

\subsubsection{Reuse-Aware Placement Policy}
\label{sec:reuse_aware_instance_placement}
After introducing memory deduplication within a single GPU, now we discuss where deduplication should take effect in cluster scale deployment. A straightforward approach is to share duplicate memory from the local GPU or even from a remote GPU node. This approach can reduce duplicate memory copies in principle, but shared memory is accessed through NVLink or network communication instead of the local HBM. Consider an inference step that reads a shared memory region of size $S$. If the memory resides in local HBM, the access time can be approximated as
\begin{equation}
T_{\mathrm{local}}(S) = \frac{S}{B_{\mathrm{HBM}}} + L_{\mathrm{local}}
\approx
\frac{S}{B_{\mathrm{HBM}}},
\end{equation}
where $B_{\mathrm{HBM}}$ is the local HBM bandwidth and $L_{\mathrm{local}}$ is the local access latency. If the same data are stored on another local GPU, the access time becomes
\begin{equation}
T_{\mathrm{nvlink}}(S) = \frac{S}{B_{\mathrm{NVLink}}} + L_{\mathrm{nvlink}},
\end{equation}
where $B_{\mathrm{NVLink}}$ is the NVLink bandwidth and $L_{\mathrm{nvlink}}$ is the additional latency of memory access. The gap between local HBM access and NVLink access can be substantial. For example, an H100 GPU provides around 3.35~TB/s of local HBM bandwidth, while the aggregate NVLink bandwidth is only about 900~GB/s. In typical GPU servers,
\begin{equation}
\begin{gathered}
B_{\mathrm{HBM}}
\gg
B_{\mathrm{NVLink}}, \\
L_{\mathrm{nvlink}}
\gg
L_{\mathrm{local}}, \\
T_{\mathrm{nvlink}}(S)
\gg
T_{\mathrm{local}}(S).
\end{gathered}
\end{equation}

The gap is more severe for the remote GPU node because the network bandwidth is much lower than NVLink. For serverless LLM inference, this overhead is not a one time cost. During decode, tokens are generated step by step. Each step repeatedly accesses the model weight and KV cache. Remote access overhead can therefore be amplified across tokens, increasing end-to-end inference latency~\cite{cachegen2024}.

For this reason, gKSM limits GPU memory deduplication to a single GPU. Duplicate memory is merged only among colocated instances on the same GPU. This single GPU sharing boundary creates the need for an instance placement policy~\cite{memorybuddies2009,muxserve2024}. A model-agnostic scheduler may place a new instance on the GPU with the most free memory or the lowest current utilization. As a result, similar instances can be spread across different GPUs, leaving duplicate model weights in memory on multiple devices. gKSM instead uses a reuse-aware instance placement policy to create the opportunities for local GPU memory deduplication. For each incoming instance $i$, gKSM decomposes its expected GPU memory demand into a shareable portion and a private portion:
\begin{equation}
M_i = S_i + P_i.
\end{equation}
The shareable portion $S_i$ includes read-only model weights and stable prefix or runtime pages~\cite{promptcache2024}. The private portion $P_i$ includes instance private KV cache, mutable runtime buffers, and memory that is not stable enough for sharing. For a candidate GPU $g$, gKSM estimates the reusable memory from resident instances as
\begin{equation}
\widehat{R}(i,g) = \mathrm{EstimateReuse}(i,\mathcal{I}_g),
\end{equation}
where $\mathcal{I}_g$ is the set of instances already placed on the GPU $g$. The incremental GPU memory cost is
\begin{equation}
\Delta M(i,g) = S_i + P_i - \widehat{R}(i,g).
\end{equation}
gKSM first filters out GPUs that do not have enough memory or compute headroom, and then selects
\begin{equation}
g^{*} = \arg\min_{g\in\mathcal{G}_i} \Delta M(i,g)
\end{equation}
among the remaining candidates. If multiple GPUs have similar incremental memory cost, gKSM prefers the GPU with more compute headroom to avoid concentrating too many active requests on one GPU. This placement policy operates at the scheduling layer and is decoupled from the memory deduplication at the driver layer. After placement, gKSM still relies on page hashes and exact content comparison before remapping duplicate pages to a shared physical handle. The same rule also supports framework transparency and partial sharing. For example, instances using vLLM and SGLang can still share duplicate GPU memory even though their batching and scheduling states are independent. Similarly, fine-tuned models with partially shared weights can still expose substantial duplicate GPU memory. They only need to expose duplicate GPU memory after being located on the same GPU.

\section{Evaluation}
\subsection{Evaluation Setup}
\noindent\textbf{Test Environment}. We evaluated gKSM on a GPU server with 4 NVIDIA A100 80~GB GPUs. The system runs CUDA 12.4 with NVIDIA driver version 570.148.08, serving instances with OpenAI APIs. All instances use bfloat16 weights with a maximum context length of 4096 tokens. Each benchmark client issues up to 32 concurrent requests.

\noindent\textbf{Workloads and Metrics}. The workload is generated from the ShareGPT conversation dataset~\cite{sharegpt}. Each benchmark client samples prompts from ShareGPT and submits requests according to a Poisson arrival process. We sweep the offered request rate from 0.5 to 4.0 req/s per instance. For the throughput experiment, each deployment policy runs the maximum number of instances that can be kept resident under its memory management strategy, and we report aggregate RPS of all active instances. For the latency experiment, we use the same fixed instance count across policies and report the mean time to first token (TTFT) and P99 TTFT over all completed requests. We deploy multiple serving instances with both vLLM instances and SGLang instances. These instances serve models from the Qwen3-8B~\cite{qwen3_technical_report}, Qwen2.5-7B~\cite{qwen25_technical_report}, GLM-4-9B~\cite{glm4_technical_report}, and Llama3-8B~\cite{meta_llama_3_8b_2024} families, together with instruction tuned and domain tuned models such as Qwen3-8B-Medical, Llama3-8B-Instruct, Qwen3-8B-Roleplay and Llama3-Guard-8B. The workload therefore contains identical model deployments, related model variants with partial weight overlap, and deployments across different inference frameworks.

\noindent\textbf{Implementation}. We implement gKSM by extending the native CUDA Driver API with custom logic built on top of the CUDA VMM API. CUDA Runtime calls are intercepted through symbol interposition, allowing gKSM to hook GPU memory management operations while preserving the original CUDA runtime behavior. The complete prototype contains approximately 4,000 lines of C++ and CUDA code, including a driver layer runtime library and a kernel level deduplication module. For serving experiments, the background scanner uses a base interval of 1~s and a maximum interval of 16~s to balance the frequency of stability checks against the scanning overhead.

\begin{figure}[t]
\centering
\begin{subfigure}{0.48\linewidth}
    \includegraphics[width=\linewidth]{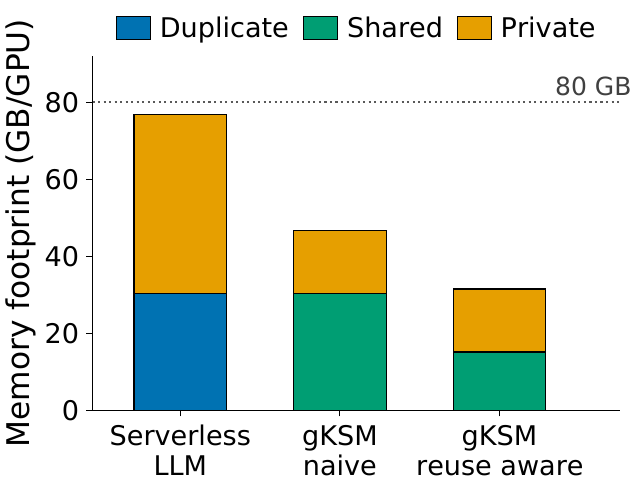}
\end{subfigure}
\hfill
\begin{subfigure}{0.48\linewidth}
    \includegraphics[width=\linewidth]{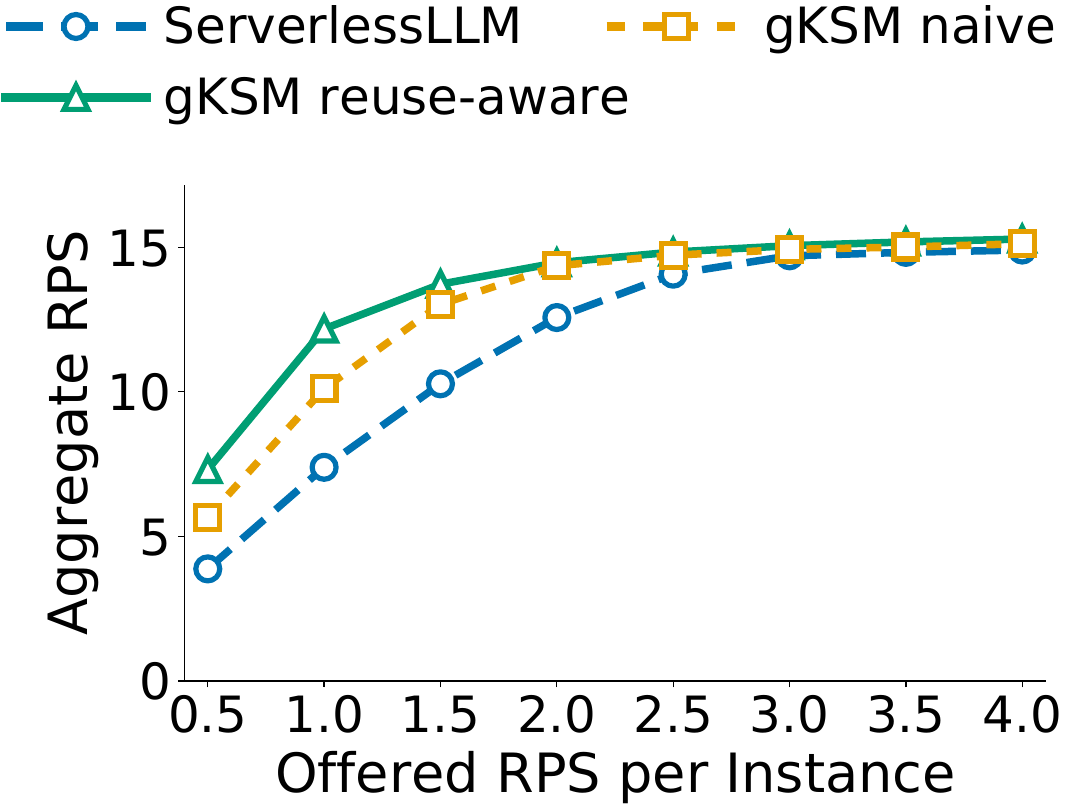}
\end{subfigure}
\caption{Memory usage and throughput in 2GPU config}
\label{fig:2gpu_sharegpt}
\end{figure}

\begin{figure}[t]
\centering
\begin{subfigure}{0.48\linewidth}
    \includegraphics[width=\linewidth]{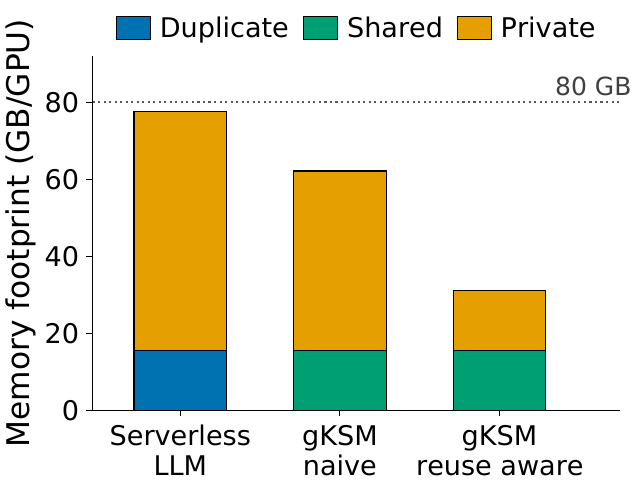}
\end{subfigure}
\hfill
\begin{subfigure}{0.48\linewidth}
    \includegraphics[width=\linewidth]{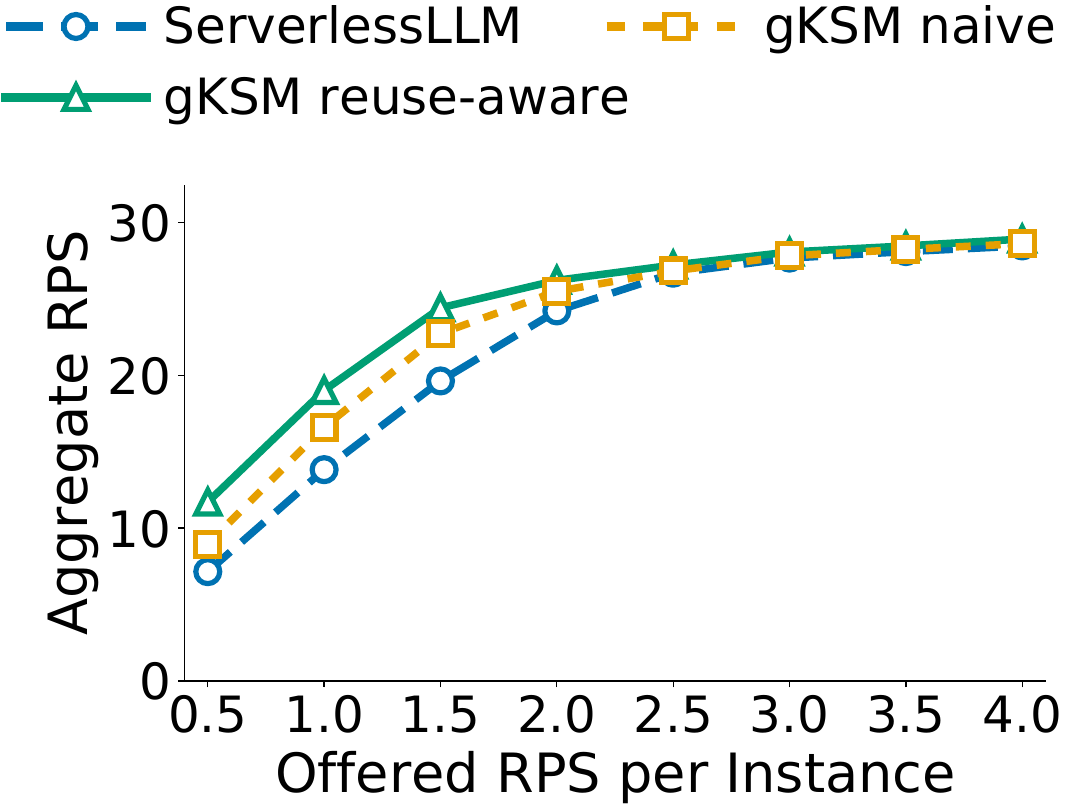}
\end{subfigure}
\caption{Memory usage and throughput in 4GPU config}
\label{fig:4gpu_sharegpt}
\end{figure}

\noindent\textbf{Deployment Policies}. We compare three deployment policies. \emph{ServerlessLLM} uses independent serving instances with separate physical copies of GPU memory. We place four instances on each GPU with different frameworks and model variants. \emph{gKSM naive} keeps the same placement policy as ServerlessLLM but enables driver layer GPU memory deduplication, enabling transparent memory sharing. \emph{gKSM reuse-aware} also enables gKSM but preferentially locates instances from the same model family on the same GPU. This placement exposes more duplicate GPU memory across frameworks and related model variants, increasing the sharing opportunities available to gKSM.

\subsection{Overall Performance}
\subsubsection{\textbf{Serverless Inference Scenarios.}} Figure~\ref{fig:2gpu_sharegpt} and~\ref{fig:4gpu_sharegpt} show the memory footprint and the actual RPS under the ShareGPT request rate sweep. In the 2 GPU setting, gKSM clearly improves aggregate throughput under low workload, where gKSM naive and gKSM reuse-aware improve throughput by about 30.8\% and 51.3\% on average over ServerlessLLM. The same trend appears in the 4 GPU setting, where the average gains are about 17.4\% and 30.7\%, respectively. This is because memory deduplication allows each GPU to hold more serving instances at the same time. As the offered load increases, the throughput gap of all system approach narrows, indicating that GPU compute becomes the bottleneck in both the 2 GPU and 4 GPU configurations. These results show that gKSM removes duplicate GPU memory copies and improves throughput while maintaining comparable serving performance at saturation.

\begin{figure}[t]
\centering
\begin{subfigure}{0.48\linewidth}
    \includegraphics[width=\linewidth]{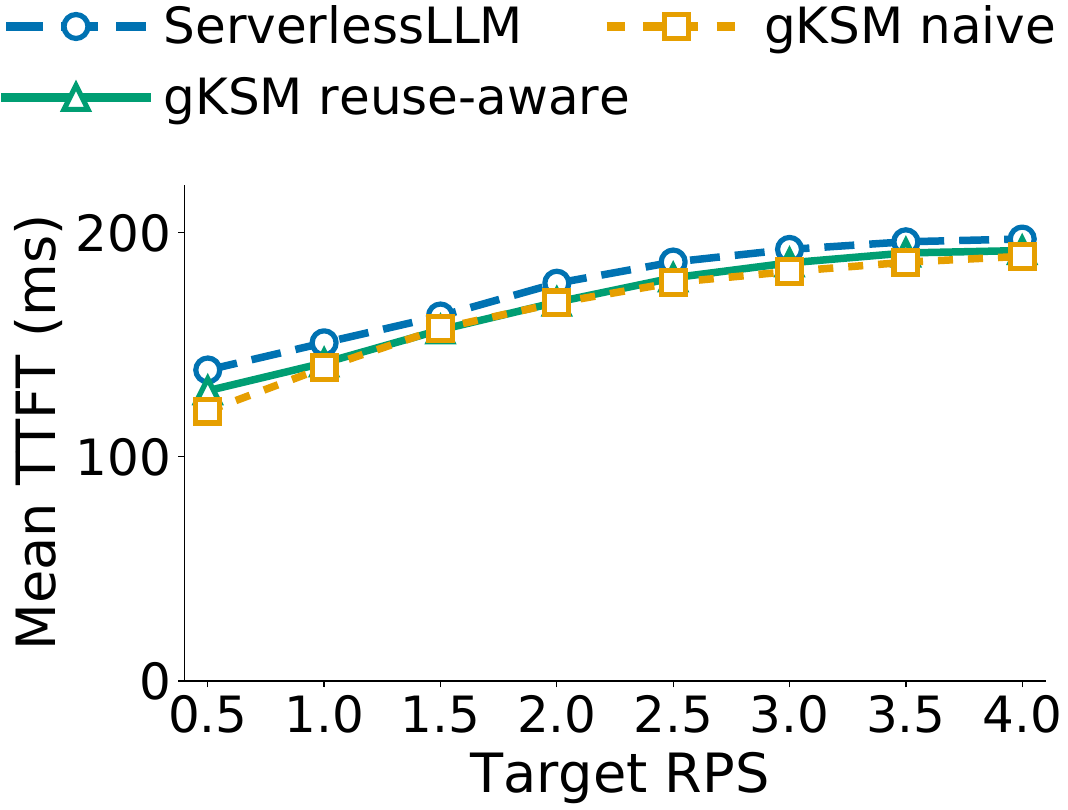}
\end{subfigure}
\hfill
\begin{subfigure}{0.48\linewidth}
    \includegraphics[width=\linewidth]{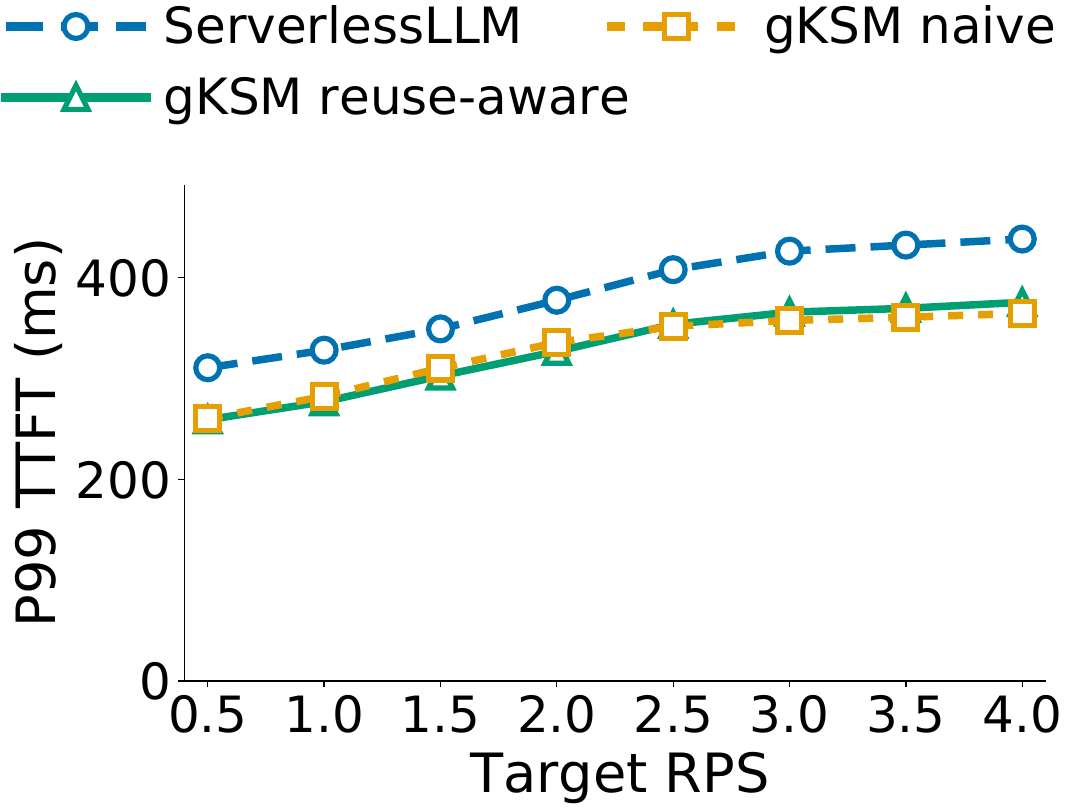}
\end{subfigure}
\caption{Mean and P99 TTFT in 2GPU config}
\label{fig:2gpu_sharegpt_ttft}
\end{figure}

\begin{figure}[t]
\centering
\begin{subfigure}{0.48\linewidth}
    \includegraphics[width=\linewidth]{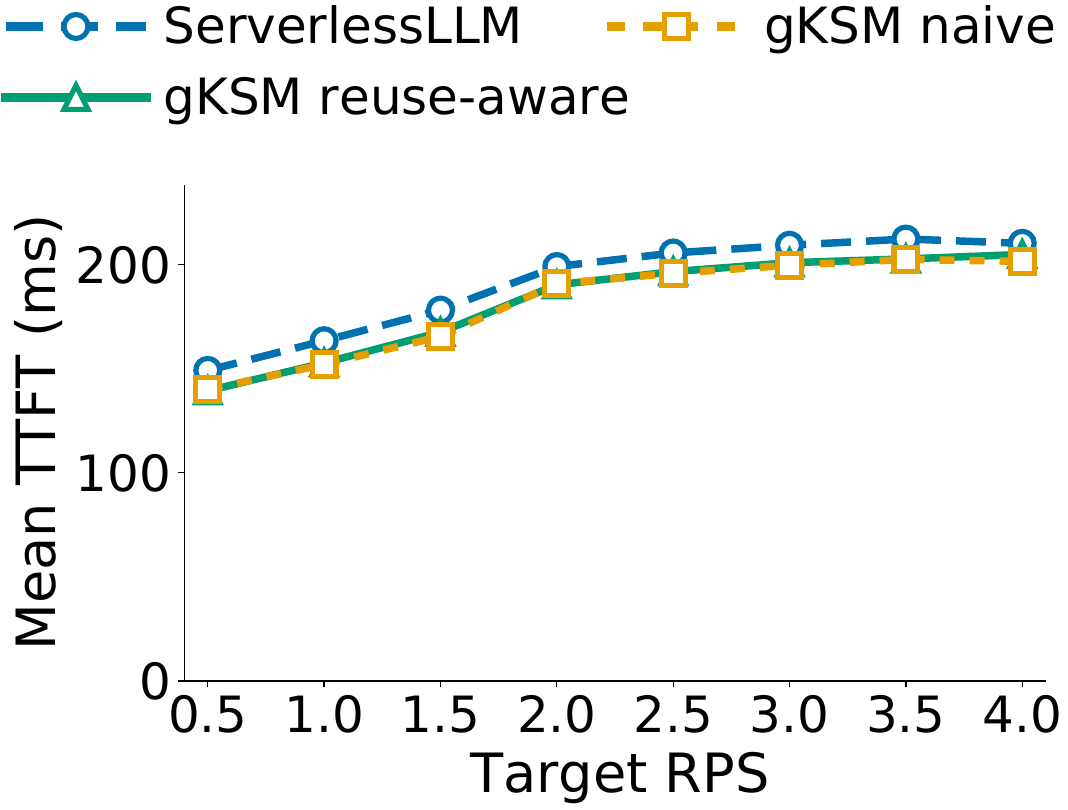}
\end{subfigure}
\hfill
\begin{subfigure}{0.48\linewidth}
    \includegraphics[width=\linewidth]{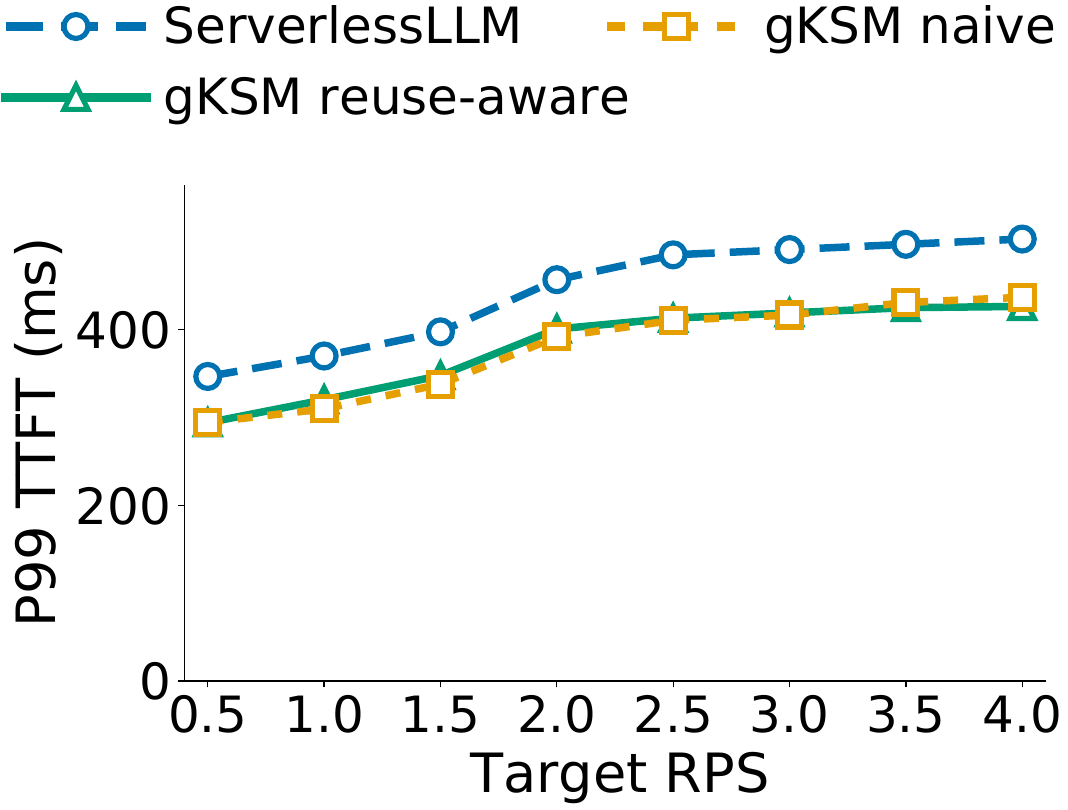}
\end{subfigure}
\caption{Mean and P99 TTFT in 4GPU config}
\label{fig:4gpu_sharegpt_ttft}
\end{figure}

Figure~\ref{fig:2gpu_sharegpt_ttft} and Figure~\ref{fig:4gpu_sharegpt_ttft} show that gKSM provides consistent TTFT improvements in different request rates. Across all request rates, gKSM naive reduces mean TTFT by 9.93 ms (5.9\%) relative to ServerlessLLM, while gKSM reuse-aware reduces mean TTFT by 7.00 ms (4.2\%). The improvement is more significant for tail latency. gKSM naive and gKSM reuse-aware reduce P99 TTFT by 55.9 ms (14.5\%) and 55.0 ms (14.4\%) on average. Under the highest request rate, ServerlessLLM reaches 438.4 ms P99 TTFT, while gKSM naive and gKSM reuse-aware reduce P99 TTFT to 364.7 ms and 375.7 ms, corresponding to reductions of 73.7 ms (16.8\%) and 62.7 ms (14.3\%), respectively. These results suggest that GPU memory deduplication reduces memory pressure and leaves more GPU memory available for KV cache allocation, improving TTFT at only a small throughput cost.

Memory usage in Figure~\ref{fig:2gpu_sharegpt} and Figure~\ref{fig:4gpu_sharegpt} reports the average memory usage per GPU under the three deployment policies. The bars decompose the memory into duplicate, shared, and private memory. ServerlessLLM does not perform deduplication, so repeated model copies appear as duplicate memory and the total usage reaches about 96\% to 97\% of GPU capacity. In the 2 GPU setting, gKSM naive and gKSM reuse-aware reduce memory usage by 39.3\% and 59.0\%, respectively. In the 4 GPU setting, the reductions are 19.9\% and 59.8\%. gKSM naive removes duplicate copies on each GPU, but its mixed placement leaves several unrelated model bases on the same GPU, which remain private and limit memory savings. gKSM reuse-aware locates instances from the same model family, so each GPU maintains a single shared model base and greatly reduces private model memory.

\begin{figure}[t]
\centering
\begin{subfigure}{0.48\linewidth}
    \includegraphics[width=\linewidth]{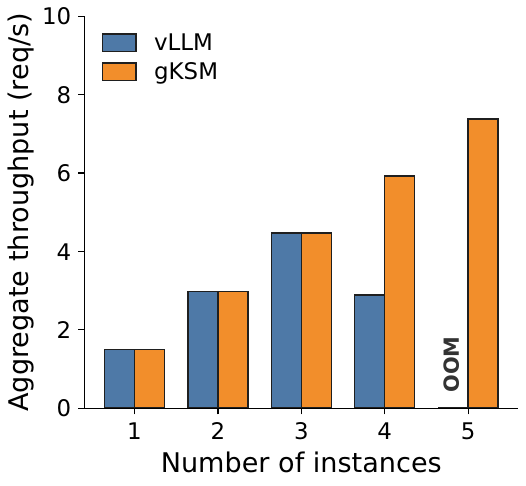}
\end{subfigure}
\hfill
\begin{subfigure}{0.48\linewidth}
    \includegraphics[width=\linewidth]{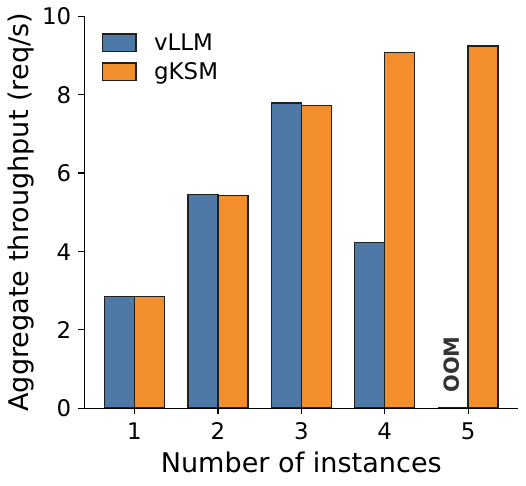}
\end{subfigure}
\caption{The aggregate throughput in memory constrained scenarios of Llama3-8B on a single A100 GPU}
\label{fig:sum_rps}
\end{figure}

\begin{figure}[t]
\centering
\begin{subfigure}{0.48\linewidth}
    \includegraphics[width=\linewidth]{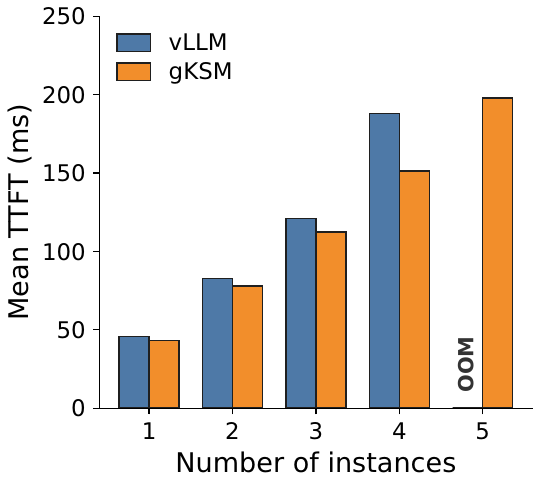}
\end{subfigure}
\hfill
\begin{subfigure}{0.48\linewidth}
    \includegraphics[width=\linewidth]{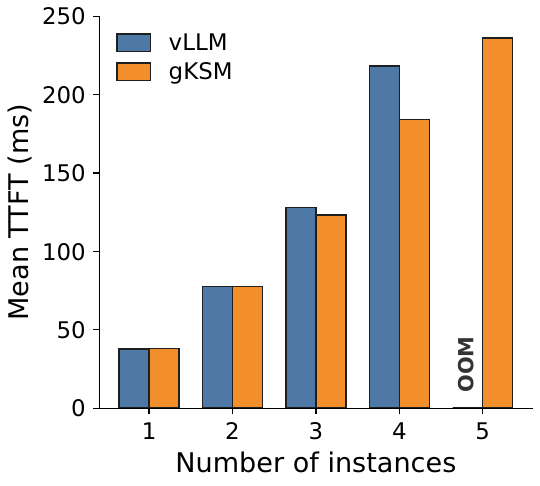}
\end{subfigure}
\caption{The mean TTFT in memory constrained scenarios of Llama3-8B on a single A100 GPU}
\label{fig:mean_ttft}
\end{figure}

\subsubsection{\textbf{Memory Constrained Scenarios.}}
\begin{figure}[t]
\centering
\begin{subfigure}{0.48\linewidth}
    \includegraphics[width=\linewidth]{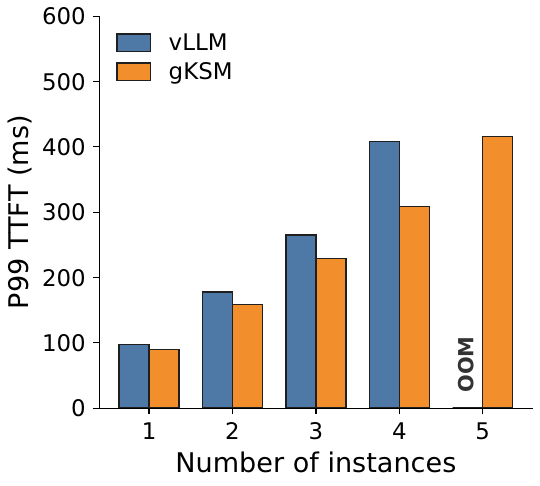}
\end{subfigure}
\hfill
\begin{subfigure}{0.48\linewidth}
    \includegraphics[width=\linewidth]{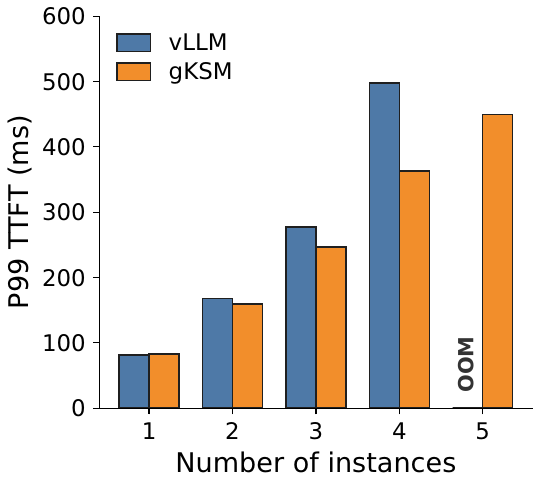}
\end{subfigure}
\caption{The P99 TTFT in memory constrained scenarios of Llama3-8B on a single A100 GPU}
\label{fig:p99_ttft}
\end{figure}

Next, we evaluated whether gKSM improves multi-instance serving capacity when GPU memory becomes the limiting resource. We deploy multiple identical Llama3-8B vLLM instances on a single NVIDIA A100 GPU and replay ShareGPT requests to each instance at request rates of 1.5 and 3 requests/s per instance. The baseline launches independent vLLM instances and falls back to CPU offloading once GPU memory is insufficient. A deployment is marked as OOM when the requested number of instances cannot be initialized or served. gKSM runs the same workload but shares duplicate GPU pages after initialization, so additional instances do not require another copy of the model weights. Figures~\ref{fig:sum_rps}, \ref{fig:mean_ttft}, and \ref{fig:p99_ttft} report aggregate throughput, mean TTFT, and P99 TTFT, respectively.

With up to three instances, both vLLM and gKSM fit within GPU memory and achieve similar throughput, showing that gKSM introduces negligible overhead before GPU memory becomes constrained. The gap emerges at four instances when the baseline begins CPU offloading due to insufficient GPU memory. Under 1.5 req/s per instance, vLLM throughput drops to 2.89 req/s, while gKSM reaches 5.92 req/s. Under 3 req/s per instance, vLLM reaches only 4.22 req/s, while gKSM reaches 9.07 req/s. The throughput degradation of the baseline is mainly caused by the reduced KV cache capacity under memory pressure, which triggers KV cache offloading and frequent CPU memory transfers, significantly slowing down the affected instances and reducing the overall request throughput. gKSM improves aggregate throughput by 2.05$\times$ and 2.15$\times$ under 1.5 req/s and 3 req/s, respectively. With five instances, the baseline fails with OOM, whereas gKSM continues serving all instances and achieves 7.38 and 9.24 req/s.

The TTFT results further show that gKSM avoids the latency overhead caused by CPU offloading under GPU memory pressure. Under the 1.5 req/s workload, gKSM reduces mean TTFT from 187.88 ms to 151.19 ms in four instances. Under the 3 req/s workload, the mean TTFT decreases from 218.13 ms to 184.24 ms. It also reduces P99 TTFT by 24.2\% and 27.0\% under the 1.5 req/s and the 3 req/s workloads, respectively. By eliminating duplicate model weights from GPU memory, gKSM allows more instances to fully reside on GPU memory and avoids the additional memory transfer overhead introduced by offloading. At the highest workload, throughput no longer scales linearly with five instances because the computation and scheduling overhead become the primary bottlenecks. However, gKSM continues to serve all instances without OOM failures and maintains high aggregate throughput.

\subsection{Performance Breakdown}
We further characterize gKSM's deduplication performance and memory savings by profiling individual CUDA memory operations, measuring COW latency and throughput under concurrent writes, and breaking down deduplicated memory into different types, analyzing how different memory components contribute to its memory savings.

\noindent\textbf{API-Level Overhead}. As shown in table~\ref{tab:apis}, the overhead of gKSM comes mainly from page hashing, VMM remapping operations, and COW handling. Among them, page hashing is the most frequent operation, but batching reduces the average hashing cost from 200.15 $\mu$s to 118.67 $\mu$s per page. The core VMM operations are lightweight, with \texttt{cuMemCreate()} and \texttt{cuMemMap()} taking only 23.62 $\mu$s and 8.16 $\mu$s on average, respectively. Although \texttt{cuMemSetAccess()} and page fault handling have higher latency per call, they are only triggered when sharing relationships or page permissions change. COW events are also rare because gKSM primarily shares read-only model weights and stable runtime pages. Once pages are shared, normal memory accesses proceed through the original virtual addresses without additional processing. As a result, these costs have only a small impact on the overall inference throughput.

\begin{table}[t]
  \centering
  \setlength{\tabcolsep}{2.5pt}
  \renewcommand{\arraystretch}{0.98}
  \begin{tabular}{@{}lrrr@{}}
    \toprule
    API & calls & total time(ms) & avg time($\mu$s) \\
    \midrule
    \texttt{hash\_xxh3(single)}  & 7798   & 1560.80  & 200.15 \\
    \texttt{hash\_xxh3(batch)}   & 160256 & 19017.35 & 118.67 \\
    \texttt{pagefault\_handle}   & 190    & 107.44   & 565.44 \\
    \texttt{cuMemCreate}         & 18592  & 439.13   & 23.62 \\
    \texttt{cuMemMap}            & 18706  & 152.62   & 8.16 \\
    \texttt{cuMemSetAccess}      & 7960   & 2358.74  & 296.32 \\
    \texttt{cuMemExport}         & 7658   & 769.82   & 100.53 \\
    \texttt{cuMemImport}         & 456    & 32.84    & 72.02 \\
    \bottomrule
  \end{tabular}
  \caption{Overhead of CUDA APIs}
  \label{tab:apis}
\end{table}

\begin{figure}[t]
\centering
\begin{subfigure}{0.49\linewidth}
    \includegraphics[width=\linewidth]{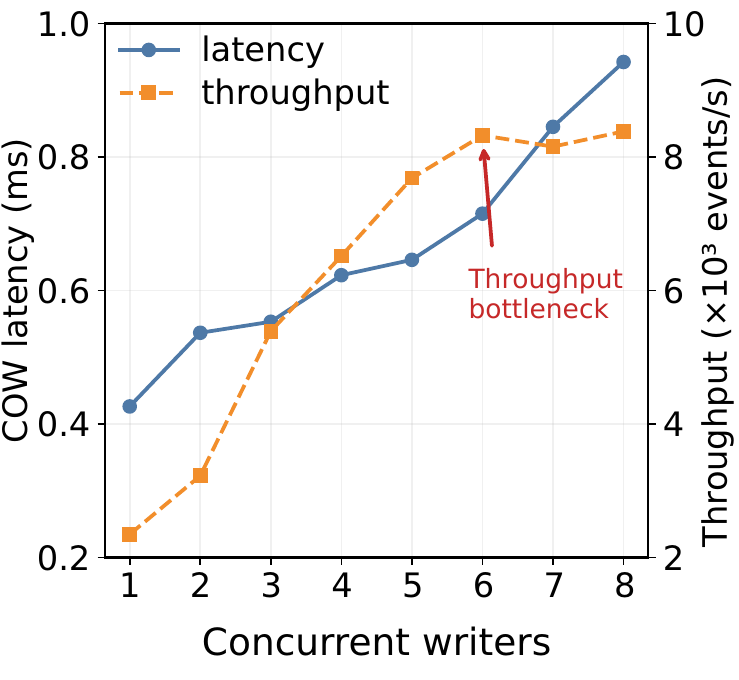}
    \caption{COW performance.}
    \label{fig:cow-stress}
\end{subfigure}
\hfill
\begin{subfigure}{0.49\linewidth}
    \includegraphics[width=\linewidth]{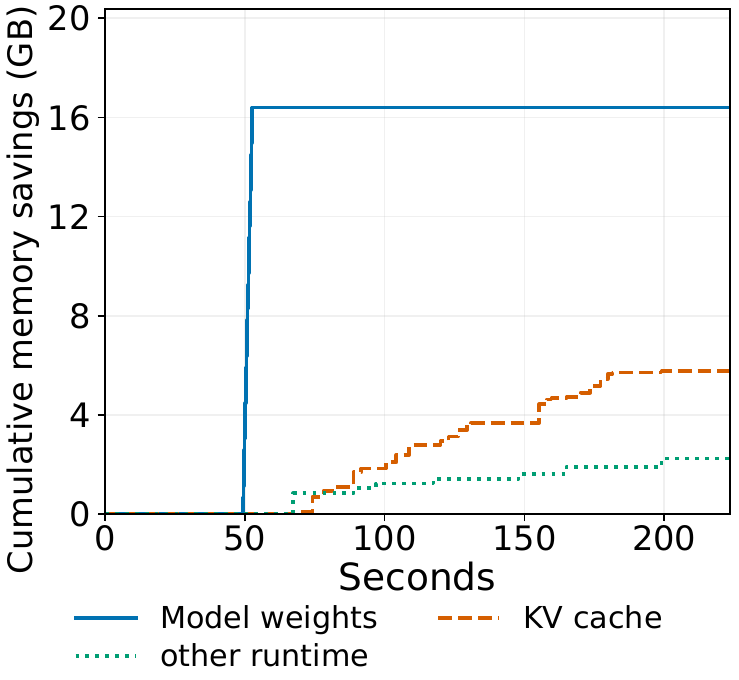}
    \caption{Cumulative memory savings.}
    \label{fig:deduplication_breakdown}
\end{subfigure}
\caption{Deduplication component analysis.}
\end{figure}

\noindent\textbf{Concurrent COW Performance}. The latency of handling cross-privilege COW fault includes kernel-to-user notification, metadata and ownership updates, private-page allocation and copying, VMM remapping, and completion acknowledgment. To further characterize COW overhead under concurrent write pressure, we synchronize 1--8 writer processes of separate CUDA streams, each performing its first write to a 2 MB page that was previously deduplicated with a read-only holder process. Figure~\ref{fig:cow-stress} reports the average COW latency and the aggregate throughput of COW handling. As shown in Figure~\ref{fig:cow-stress}, the average COW latency increases from 0.426 ms with one writer to 0.942 ms with eight writers, while the aggregate throughput increases from 2346.2 to 8382.2 events/s, a $3.57\times$ increase. At 6--8 concurrent writers, throughput plateaus at about 8300 events/s while latency continues to rise. Under such high concurrency, lock contention in the COW handling path and blocking waits in the page fault queue become the primary bottlenecks, preventing throughput from scaling while increasing the COW latency. Given that the colocation of four instances per-GPU represents the practical configuration in our evaluation, these results show that gKSM maintains COW performance for practical deployment.

\noindent\textbf{Memory Deduplication Breakdown}. We classify deduplicated pages as model weights, KV cache, and other runtime data by logging GPU address ranges of framework-managed memory regions, mapping each page to its corresponding region, and aggregating deduplicated bytes by memory type. Figure~\ref{fig:deduplication_breakdown} illustrates their cumulative contributions over time of two instances configured with the Qwen3-8B model. Model weights are handled by zero-copy loading through the intercepted \texttt{cuMemcpyHtoD} path. Their contribution rises rapidly during loading and then plateaus as model weights remain read-only during inference. In contrast, KV cache and other runtime data are written directly into preallocated GPU buffers by CUDA kernels and normally bypass this loading path. Background scanning identifies stable duplicate pages in these regions, producing stepwise increases as additional pages are merged. At the end of the illustrated interval, the cumulative deduplicated memory is approximately 24.4~GB, with model weights, KV cache, and other runtime data contributing 16.40~GB (67.2\%), 5.76~GB (23.6\%), and 2.24~GB (9.2\%), respectively. Weight deduplication is the largest component, while the runtime portion demonstrates that scanning provides substantial additional savings.

\subsection{Cluster Level Simulation}
We further evaluated gKSM at the cluster level by simulating and replaying an anonymized pod-level inference trace derived from Alibaba Cloud serverless workloads~\cite{yan2025Understanding} on clusters with 1 to 8 GPU nodes. Each node contains 8 A100 GPUs with 80~GB GPU memory. We compare gKSM with ServerlessLLM and SLInfer, and focus on two metrics: the fraction of pod steps that can be served and the resource balance ratio on active nodes. The resource balance ratio is defined as the average value of compute utilization divided by memory utilization across active GPUs. A higher ratio means that the cluster is less dominated by memory pressure and can use GPU compute resources more effectively.

\begin{figure}[t]
  \centering
  \includegraphics[width=0.85\linewidth]{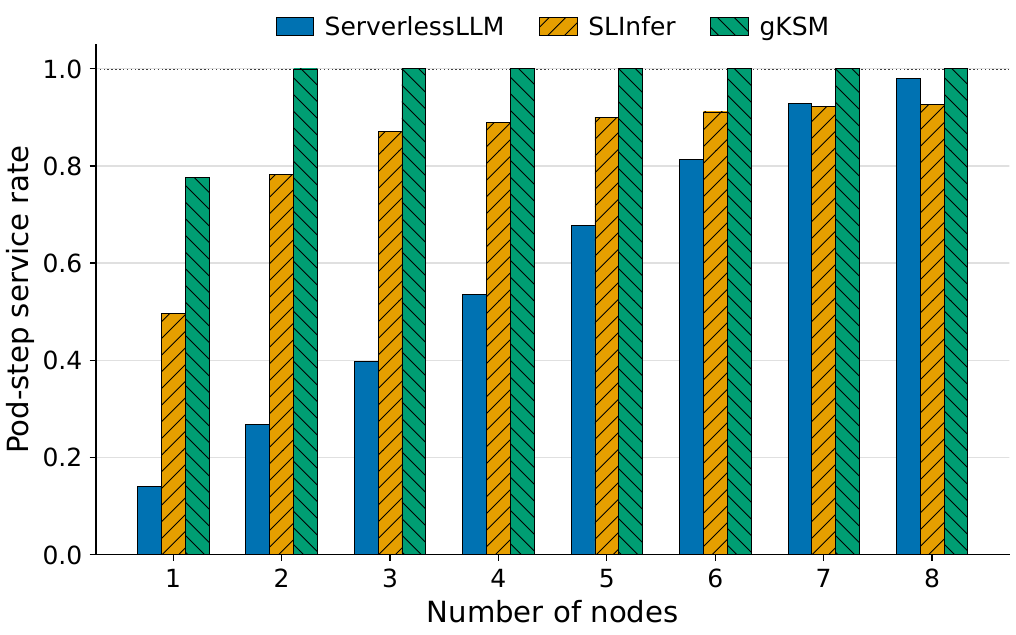}
  \caption{Inference Service Rate under Cluster Simulation}
  \label{fig:cluster_service_rate}
\end{figure}

Figure~\ref{fig:cluster_service_rate} shows the fraction of pod steps that can be served in cluster scale. With only one node, gKSM already serves 77.6\% of all pod steps, compared to 49.7\% for SLInfer and 14.0\% for ServerlessLLM. With two nodes, gKSM reaches 99.97\%, while the latter two remain at 78.2\% and 26.9\%, respectively. From three nodes onward, gKSM serves the whole workload. In contrast, ServerlessLLM still needs eight nodes to approach 98.0\%, and SLInfer reaches 100\% at eight nodes. It shows that the memory saved by deduplication can be turned directly into higher cluster capacity.

\begin{figure}[t]
  \centering
  \includegraphics[width=0.85\linewidth]{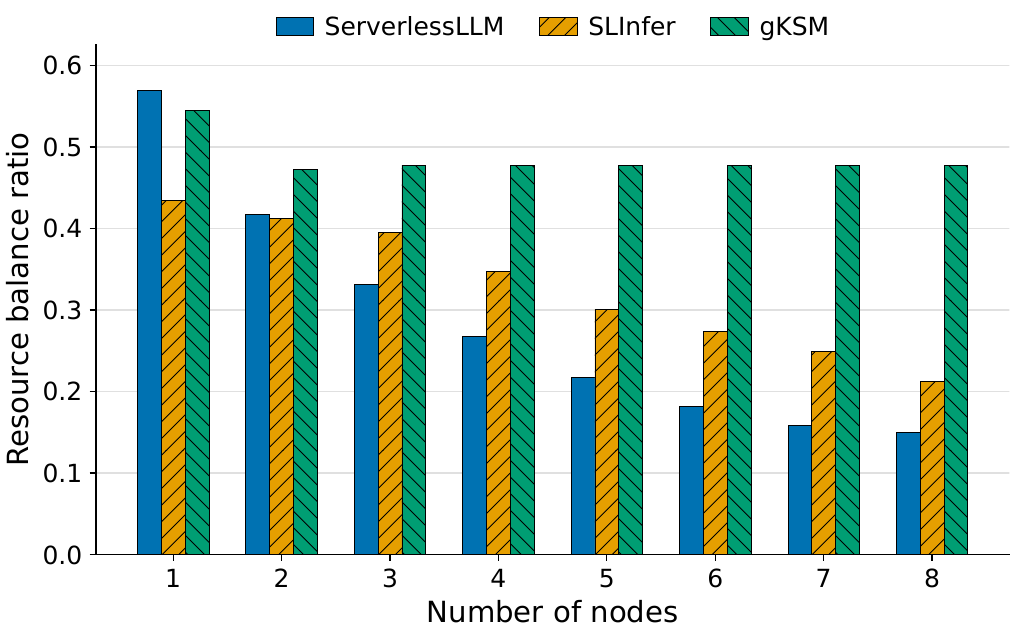}
  \caption{Resource Balance Ratio under Cluster Simulation}
  \label{fig:cluster_resource_balance}
\end{figure}

Figure~\ref{fig:cluster_resource_balance} shows the balance ratio of resources in active nodes. After two nodes, gKSM only decreases slightly from 0.472 to 0.436, which means active GPUs remain relatively balanced between compute demand and memory demand. The same ratio drops steadily for the two baselines as the cluster grows. For ServerlessLLM, it falls from 0.417 at two nodes to 0.150 at eight nodes. For SLInfer, it falls from 0.413 to 0.212 over the same range. This trend shows that the baselines spread across more GPUs, but memory remains the main bottleneck on the nodes that are still active. gKSM keeps similar instances on the same GPU and removes duplicate pages in place, so the cluster uses memory more effectively and reaches high service capacity with fewer nodes.

\section{Discussion}
\noindent\textbf{Security.}
As in other CPU memory deduplication systems, there are side-channel risks in gKSM~\cite{differenceengine2008,singleton2012,waldspurger2002memory}. However, their practical impacts are limited by the lower sensitivity in LLM inference scenarios and trust-domain restrictions. Compared with conventional CPU memory deduplication, gKSM operates at 2MB page granularity, making fine-grained probing more difficult. For public or authorized model weights, a deduplication signal may reveal only a few model weight pages, which is insufficient to reconstruct the entire model weight. Considering that KV cache retains attention states derived from prompts~\cite{kwon2023efficient,promptcache2024}, gKSM can restrict its deduplication to operator-defined trust domains.

\noindent\textbf{Deployment and Native Support.}
GPU memory deduplication can be provided by the cloud vendor or hardware provider. If NVIDIA provides GPU memory deduplication service, most parts of gKSM remain necessary, such as page management, duplication detection, stable-page scanning, and reuse-aware placement. In particular, reuse-aware placement remains relevant because sharing opportunities depend on colocation, as demonstrated in CPU virtualized systems~\cite{memorybuddies2009}. However, NVIDIA may implement COW and page merging more efficiently in its kernel.

\section{Related Work}
\noindent\textbf{Serverless Inference.}
Serverless inference has been studied as a way to elastically serve heterogeneous models under dynamic workloads. Systems such as INFaaS and AlpaServe optimize model scheduling and parallelism for inference services~\cite{infaas2021,alpaserve2023}. Recent serverless LLM systems further reduce cold start and memory pressure through model offloading, migration, and batch merging~\cite{serverlessllm2024,prism2025,slinfer2026}. These works mainly address the problem from the scheduling and serving layers. However, these systems optimize above the GPU memory layer and do not address duplicate physical memory across colocated instances. This motivates our driver-layer GPU memory deduplication design for serverless inference.

\noindent\textbf{GPU Memory Deduplication.}
Memory deduplication has a long history in CPU and virtualized systems. Related systems such as KSM and Difference Engine scan memory pages, merge identical physical frames, and rely on kernel supported COW to preserve isolation~\cite{singleton2012,differenceengine2008,miller2012ksmpp}. Recent works improve deduplication efficiency through better page grouping or scanning policies~\cite{buddy2020,memorybuddies2009,userguided2023,yang2026token}. GPU memory systems mainly focus on memory management and defragmentation to optimize KV cache management for LLM serving. However, these works do not provide direct memory deduplication between independent GPU instances~\cite{buddy2020,mosaic2017,guo2024gmlake,kwon2023efficient,vattention2025}.

\section{Conclusion}
This paper presents gKSM, a transparent GPU memory deduplication framework for serverless LLM inference. Motivated by the large amount of duplicate GPU memory across colocated model instances, gKSM enables transparent memory sharing while preserving instance isolation. Our results show that memory deduplication can substantially improve GPU memory efficiency and system scalability for serverless inference. We view gKSM as a first step toward transparent GPU memory sharing for large scale LLM serving. As model deployment continues to grow, new opportunities will emerge to exploit memory duplication in the AI world.

\begin{acks}
We sincerely thank anonymous shepherd and reviewers for their insightful suggestions. This work was supported by NSFC (62472392, 62502189) and Major Science and Technology Projects of Anhui Province (S202305a12020032). 
\end{acks}

\newpage

\setlength{\bibsep}{1pt}
\bibliographystyle{ACM-Reference-Format}
\bibliography{ref}

\end{document}